\documentclass[12pt,preprint2]{aastex6}

\usepackage{epstopdf}
\usepackage{array}
\usepackage{gensymb}
\usepackage{tablefootnote}
\usepackage{graphicx}
\usepackage{bigstrut}
\usepackage{color}

\shorttitle{\emph{Fermi} GBM Observations of V404 Cyg During its 2015 Outburst}
\shortauthors{P. Jenke et al.}

\begin{document}


\title{\emph{Fermi} GBM Observations of V404 Cyg During its 2015 Outburst}



\author{P. A. Jenke\altaffilmark{1}, C. A. Wilson-Hodge\altaffilmark{3}, Jeroen Homan\altaffilmark{4,5}, P. Veres\altaffilmark{1}, M. S. Briggs\altaffilmark{1}, E. Burns\altaffilmark{1}, V. Connaughton\altaffilmark{2}, M. H. Finger\altaffilmark{2}, M. Hui\altaffilmark{3}}
\altaffiltext{1}{University of Alabama in Huntsville, Huntsville, AL 35805, USA, peter.a.jenke@nasa.gov}
\altaffiltext{2}{Universities Space Research Association, Huntsville, AL 35805, USA.}
\altaffiltext{3}{Marshall Space Flight Center, Huntsville, AL 35812, USA}
\altaffiltext{4}{MIT Kavli Institute for Astrophysics and Space Research, 77 Massachusetts Avenue 37-582D, Cambridge, MA 02139, USA}
\altaffiltext{5}{SRON, Netherlands Institute for Space Research, Sorbonnelaan 2, 3584 CA Utrecht, The Netherlands}
%






\begin{abstract}
V404 Cygni was discovered in 1989 by the \emph{Ginga} X-ray satellite during its only previously observed X-ray outburst and soon after confirmed as a black hole binary.  On June 15, 2015, the Gamma Ray Burst Monitor (GBM) triggered on a new outburst of V404 Cygni.  We present 13 days of GBM observations of this outburst including Earth occultation flux measurements, spectral and temporal analysis. The Earth occultation fluxes reached 30 Crab with detected emission to 100 keV and determined, via hardness ratios, that the source was in a hard state.  At high luminosity, spectral analysis between 8 and 300 keV showed that the electron temperature decreased with increasing luminosity.  This is expected if the protons and electrons are in thermal equilibrium during an outburst with the electrons cooled by the Compton scattering of softer seed photons from the disk.  However, the implied seed photon temperatures are unusually high, suggesting a contribution from another source, such as the jet.  No evidence of state transitions is seen during this time period.  The temporal analysis reveals power spectra that can be modeled with two or three strong, broad Lorentzians, similar to the power spectra of black hole binaries in their hard state.  
\end{abstract}





\section{Introduction}
V404 Cygni, hereafter V404 Cyg, was first identified as an X-ray transient with the \emph{Ginga} satellite during the 1989 flaring event \citep{Makino_1989}.  Using archival optical data, V404 Cyg was associated with what was thought to be a nova in two previous outbursts in 1938 and 1956.   Optical observations after the 1989 flaring event revealed an orbital ephemeris with an orbital period of 6.5 days, an inclination of 56 degrees, and most importantly a mass function of $6.08\pm0.16$ M$_\sun$ \citep{CasaresandCharles_1994} making the source one of the first confirmed black hole systems with a black hole mass $\sim$ 10 M$_{\odot}$.  A radio parallax distance of $2.39\pm0.14$  kpc \citep{Miller-Jones_2009}  allows precise estimates of the intrinsic luminosity and makes this one of the closest known black hole systems.  V404 Cyg's large separation from its companion along with optical H$\alpha$ observations indicate that the system develops a very large accretion disk which contains an inordinate amount of material ensuring dramatic flares when the inner region of the accretion disk breaks down \citep{Remillard06}.

At 18:31:38 UT on 2015, June 15, the \emph{Swift} Burst Alert Telescope (BAT) triggered and located V404 Cyg \citep{Barthelmy_2015}.  Twenty eight minutes later, the \emph{Fermi} Gamma Ray Burst Monitor (GBM) triggered on an X-ray source with a subsequent ground localization consistent with V404 Cyg.  
\section{Gamma Ray Burst Monitor}
	 GBM is an all sky monitor whose primary objective is to extend the energy range over which gamma-ray bursts are observed in the Large Area Telescope (LAT) on \emph{Fermi} \citep{Meegan_2009}.  GBM consists of 12 NaI detectors with a diameter of 12.7 cm and a thickness of 1.27 cm and two BGO detectors with a diameter and thickness of 12.7 cm.  The NaI detectors have an energy range from 8 keV to 1 MeV while the BGOs extend the energy range to 40 MeV.  GBM has three continuous data types: CTIME data with nominal 0.256-second time resolution and 8-channel spectral resolution used for event detection and localization, CSPEC data with nominal 4.096-second time resolution  and 128-channel spectral resolution which is used for spectral modeling, and Continuous Time Tagged Event (CTTE) data which has a timing precision of $2\mu$s.  All three data types are utilized in the following analysis. 
	 
	The GBM flight software was designed so that GBM can trigger on-board in response to impulsive events, when the count rates recorded in two or more NaI detectors significantly exceed the background count rate on at least one time-scale from 16 ms to 4.096 s in at least one of four energy ranges above 25 keV.  The lower energy and longer time-scales are not used with the on-board triggering algorithms owing to strong variations in background rates that are incompatible with a simple background modeling needed for automated operation on a spacecraft.
	 

\section{Observations}
GBM triggered on V404 Cyg 169 times between June 15-27.  The source reached a brightness of 30 Crab with emission to 300 keV.  With an 8 sr field of view, GBM was able to observe the entire outburst with a duty cycle of 57\%.   GBM is not an imaging instrument but uses the differential response of its 12 NaI detectors to locate sources to a few degrees \citep{Connaughton_2015}.  Even though localization is possible, GBM's high background rates can make analysis challenging.   To mitigate this limitation, three techniques are employed  to analyze this source: the Earth occultation technique, choosing times of high signal to noise such as GBM triggered events, and using GBM's 2$\mu$s timing precision to extract temporal information from the data.  These techniques and their results are detailed below.
\subsection{Earth Occultation Observations}
\begin{figure}[h!]
\begin{centering}
\includegraphics[width=0.9\columnwidth]{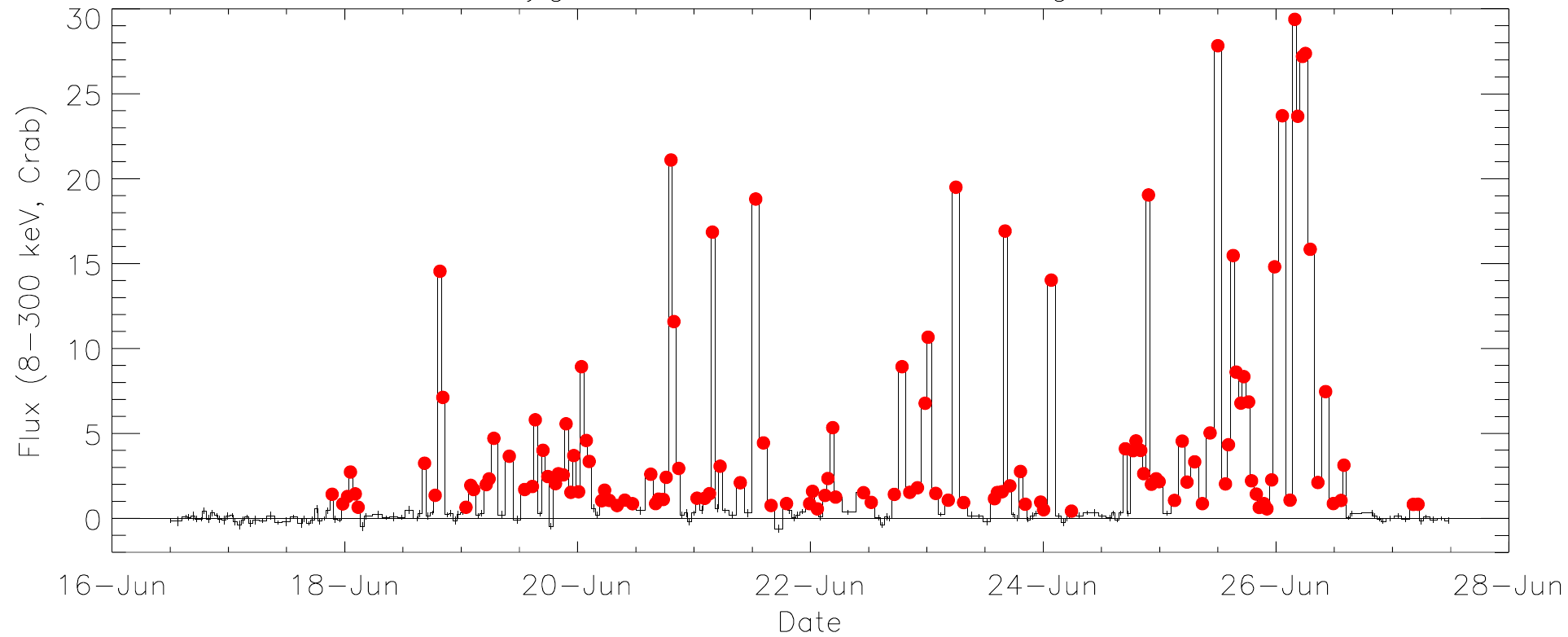}
\caption{V404 Cyg light curve measured with Fermi GBM in the 8-300 keV band. Fluxes are normalized to GBM measurements of the Crab in the same band. Red circles indicate 3 $\sigma$ or better detections in a single occultation step.}
\label{fig:occ_lc}
\end{centering}
\end{figure}

The Earth Occultation software, described in detail in \cite{WilsonHodge2012}, fits the GBM CTIME data with a quadratic background plus models of occultation steps for the source of interest and any other bright sources occulting during the 4-minute fit window. The occultation step models incorporate atmospheric transmission and an assumed source flux model for each source in the fit. Independent fits are performed for each detector and each energy channel. For V404 Cyg, the assumed flux model was based on INTEGRAL SPI measurements \citep{Rodriguez15}. Steps during solar flares, and when the constant background term was more than 3$\sigma$ away from its Gaussian mean from pre-outburst data 2008 August to 2015 June 16 were removed from the analysis. High values of the constant term correlate with periods of high KP index (high particle activity). Figure~\ref{fig:occ_lc} shows the light curve based on single occultation steps for V404 Cyg in the 8-300 keV band (GBM CTIME channels 0-4). V404 Cyg flux measurements in each energy channel were normalized to the average flux for the Crab nebula and pulsar for the time period 2015 June 17-July 2.  Significant detections of a source within a single occultation step with GBM is unusual and is an indication of how bright V404 Cyg's flares were.

 \citet{Remillard06} described ways of defining black hole states, including a method based on radio properties, X-ray power density spectra, and hardness intensity diagrams. Hardness ratios were defined as 8.6-18.0 keV counts/5.0-8.6 keV counts measured with the {\it Rossi X-ray Timing Explorer (RXTE)} Proportional Counter Array, for which the Crab nebula yielded HR=0.68. \citet{Remillard06} found that sources with HR$>$0.68, harder than the Crab Nebula, corresponded to the hard state in both \citet{McClintock06} model and in the unified jet model \citep{Fender04} and HR$<$0.2 corresponded to the steep power law state. Further discussion in \citet{Remillard06}, and references therein, points out that in the hard state, an exponential cutoff near 100 keV is often found, while QPOs may or may not be present. \citet{Remillard06} emphasize that luminosity is not a criterion for identifying X-ray states in either prescription. 

\begin{figure}[h!]
\begin{centering}
\includegraphics[width=0.9\columnwidth]{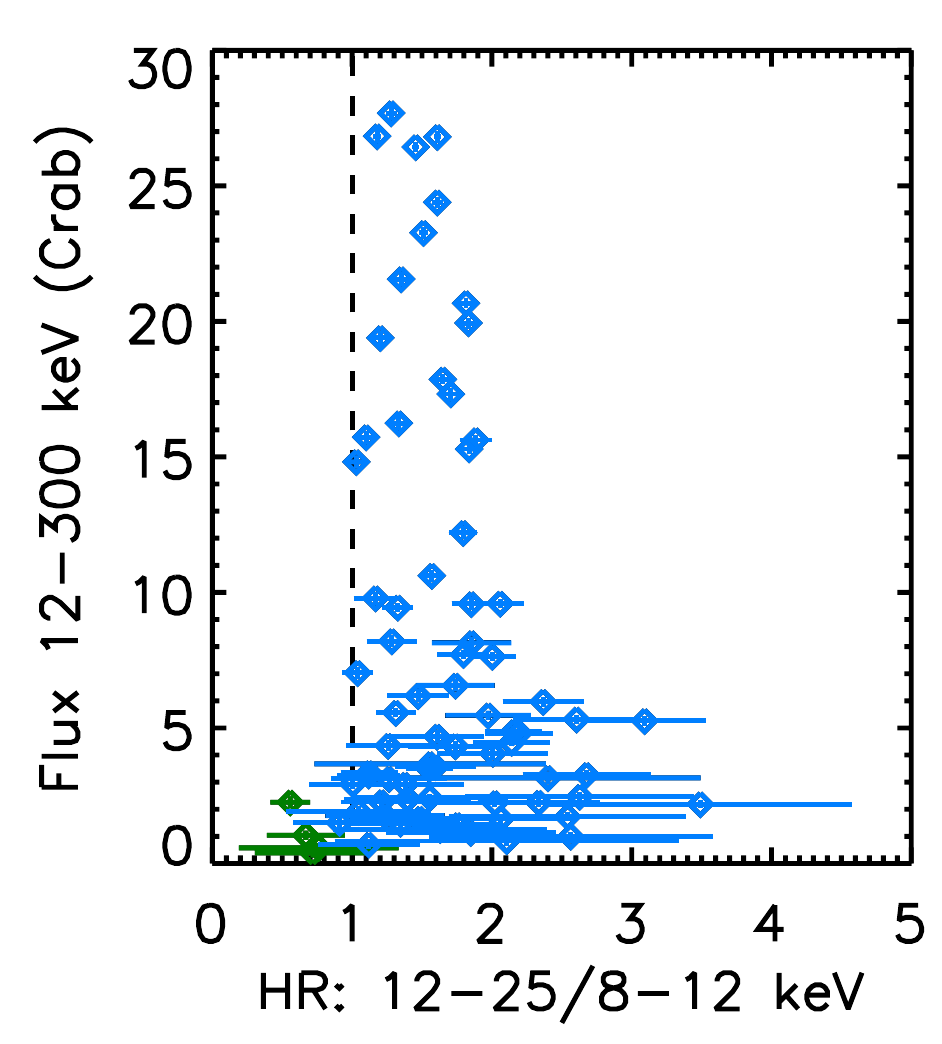}
\caption{V404 Cyg hardness ratio (12-25 keV flux/8-12 keV flux) vs total flux in the 12-300 keV range. The dashed vertical line indicates the Crab hardness ratio for these bands. It is at 1.0 because the fluxes are normalized to the Crab. Blue diamonds indicate where V404 Cyg is harder than the Crab and green diamonds indicate where V404 Cyg is softer than the Crab.}
\label{fig:occ_hid}
\end{centering}
\end{figure}

To compare GBM measurements of V404 Cyg to these studies, hardness ratios were generated by dividing the single step flux measured in the 12-25 keV band by the flux in the 8-12 keV band, the lowest two bands available in GBM data and closest to the canonical {\it RXTE} bands, shown in Figure~\ref{fig:occ_hid}. The majority of the GBM hardness ratios (blue diamonds in Figure~\ref{fig:occ_hid}) are harder than the Crab, suggesting that V404 Cyg spent the majority of its outburst in the hard state even though it was emitting at a large fraction of its Eddington luminosity. 
 
\subsection{Spectral Analysis}
During the hard state the BH disk is truncated and the inner region is filled with a hot ($>50$ keV) advection-dominated optically thin accretion flow typically referred to as the corona.  There are many unanswered questions regarding the corona including its size, location, and shape but it is generally agreed that the high energy emission originates from the hot electrons upscattering soft photons from the accretion disk.  We performed spectral analysis of the hard X-ray emission between 8 and 1000 keV in order to better understand this hot Comptonized corona in the vicinity of the black hole.
\subsubsection{Data Selection}
2000s  of GBM CSPEC data centered on the trigger times for all 169 triggers were selected for spectral analysis.  Detectors with angles between the source and detector bore sight greater than 60 degrees were excluded.  Using RMFIT, a forward-folding spectral analysis software often used in GBM gamma ray burst studies\footnote{http://fermi.gsfc.nasa.gov/ssc/data/analysis/rmfit/}, a polynomial background was fit to each detector in each energy channel between 8 and 1000 keV using times before and after the flare.  Times around the trigger times, when a flare was evident, were chosen and response matrices were created from a response model constructed from simulations incorporating the Fermi spacecraft mass model into GEANT4 \citep{GEANT4}.  Source selection was limited to 200 s in order to ensure that an adequate background model could be fit and that the spacecraft response would not significantly change during the times of data selection.  The background subtracted data was fit using RMFIT to the CompST model \citep{Sunyaev_Titarchuk_1980} and the residuals in each detector were compared for consistency.  Consistent residuals across detectors are an indication that the background selection is reasonable.  Triggered events in which a good background model could not be constructed were rejected.  For the remaining 155 events in which an acceptable spectral fit was possible,  the background and the total spectrum were exported for analysis in XSPEC \citep{xspec_1996}.

\subsubsection{Integrated Spectral Analysis Results}
We chose spectral models to model the hot comptonized corona surrounding the black hole.
 The CompTT model \citep{Titarchuk_1994} and the REFLECT*CompTT  were successful in representing the data.  The REFLECT*CompTT model resulted in a reflection component that often dominated the spectrum where $\Omega/2\pi >> 1.0$.  The CompTT model alone resulted in a high seed photon temperature that averaged above 5 keV.  We fixed the seed photon temperature to 1 keV and fixed the reflection amplitude to $\Omega/2\pi = 1.0$ for the REFLECT*CompTT model.  In 130 out of 155 spectral fits we were able to reject (68\% level and 110 out of 155 at the 95\% level) the REFLECT*CompTT in favor of the CompTT model with the high seed photon index.  In addition the REFLECT*CompTT model with amplitude and seed photon temperature fixed often resulted in a high ($\overline{\tau} = 3.3$) optical depth.
 
 We also choose the CUTOFFPL model for its simplicity but the model was less successful at representing the data.  The high energy cut-off, at times, was much lower than the 100 keV often observed for stellar mass black hole systems in the hard state \citep{Malzac_2012,Remillard06} and often near GBM's low energy threshold resulting in fits with unconstrained parameters and very hard indexes.  Nevertheless, the CUTOFFPL is useful for tracing the spectral variability between triggered flares and there is no evidence for systematic softening or hardening of the spectrum during this period.  

\begin{figure}[h!]
\begin{centering}
 
\includegraphics[width=.999\columnwidth]{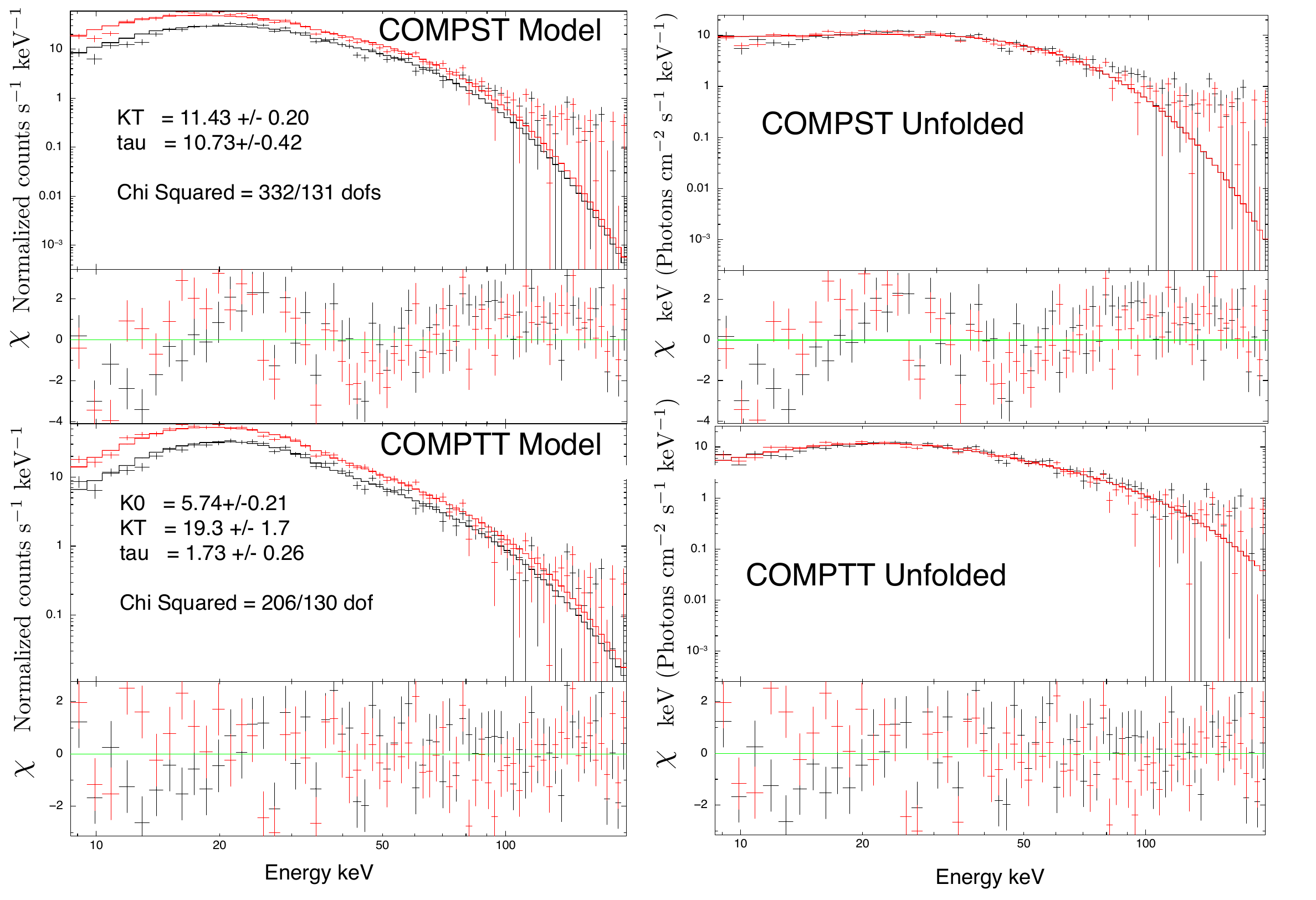}
\caption{Time integrated spectral model fitting the triggered event bn150626171 using 12 seconds of triggered data.  The figure on the top left is the spectral fit using the CompST model with residuals below.  The red and black points are the normalized counts from two of the NaI detectors while the red and black curves are the best fit model.   The figure on the bottom left is the spectral fit using the CompTT model with residuals below.  The best fit parameters are given in the figures.  The figures on the right are the same fit results as the left except shown as the unfolded spectrum (E F(E)).  The fit for the CompTT model is a significant improvement over the CompST model.}
\label{fig:CompTT}
\end{centering}
\end{figure}

The CompTT model seed photon temperature, $kT_{\rm photon}$, is a free parameter and allows us to probe emission from the up-scattering of hot seed photons as apposed to the CompST model which assumes a cold distribution of seed photons.  The CompTT model fits resulted in an average seed photon temperature of 5.9$\pm$1.3 keV consistent with INTEGRAL observations \citep{Natalucci_2015, Roques_2015}.  This high photon seed temperature is inconsistent with photons from an accretion disk ($kT_{\rm photon} < 1$ keV) and the cold photons assumed in the CompST model.  The average optical depth, $\tau$, is 1.45 $\pm$ 1.0 ($\bar{\sigma_{\tau}} = 0.56$) (see Figure \ref{fig:CompTT}) which is consistent with the moderate electron temperatures measured for this source \citep{Droulans_2010,Malzac_2012}.

%

The results for the CompTT model are summarized in Figure \ref{fig:CompTT} and detailed in the online table.  The $\chi^2$ could be improved be ignoring the energy band between 30 and 40 keV where there is a poorly modeled K-edge.  The K-edge does not affect the spectral results only the quality of the fit and removing those energy bins makes it more difficult to constrain the parameters.  There does not appear to be any evolution in the spectral parameters that would signify a state change.

\begin{figure}[h!]
\begin{centering}
\includegraphics[width=0.9\columnwidth]{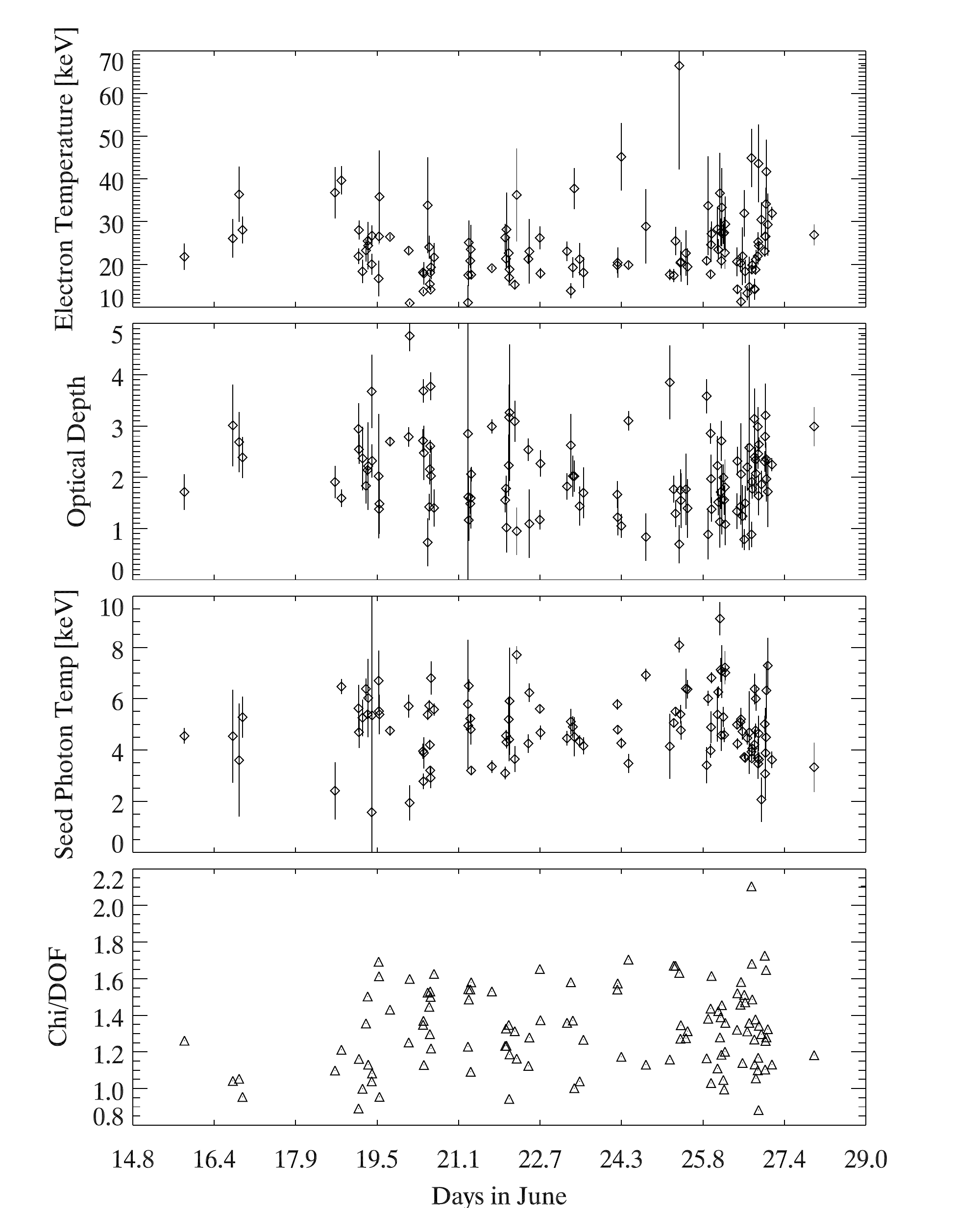}
\caption{Time integrated spectral results for each triggered event specified by days in June.  Only the results where the electron temperature can be constrained are shown.  The top panel is the electron temperature which generally varies between 10-40 keV.  The panel below is the optical depth which varies between 1 and 4.  The third panel is the seed photon temperature which varies between 2 and 8 keV.  The last panel shows the reduced $\chi^2$ for the fit.  For GBM data, anything below 1.6 is considered a successful fit.}
\label{fig:spectral_results}
\end{centering}
\end{figure}

\subsubsection{Time Resolved Spectral Analysis Results}
\begin{figure}[h!]
\begin{centering}
\includegraphics[width=0.9\columnwidth]{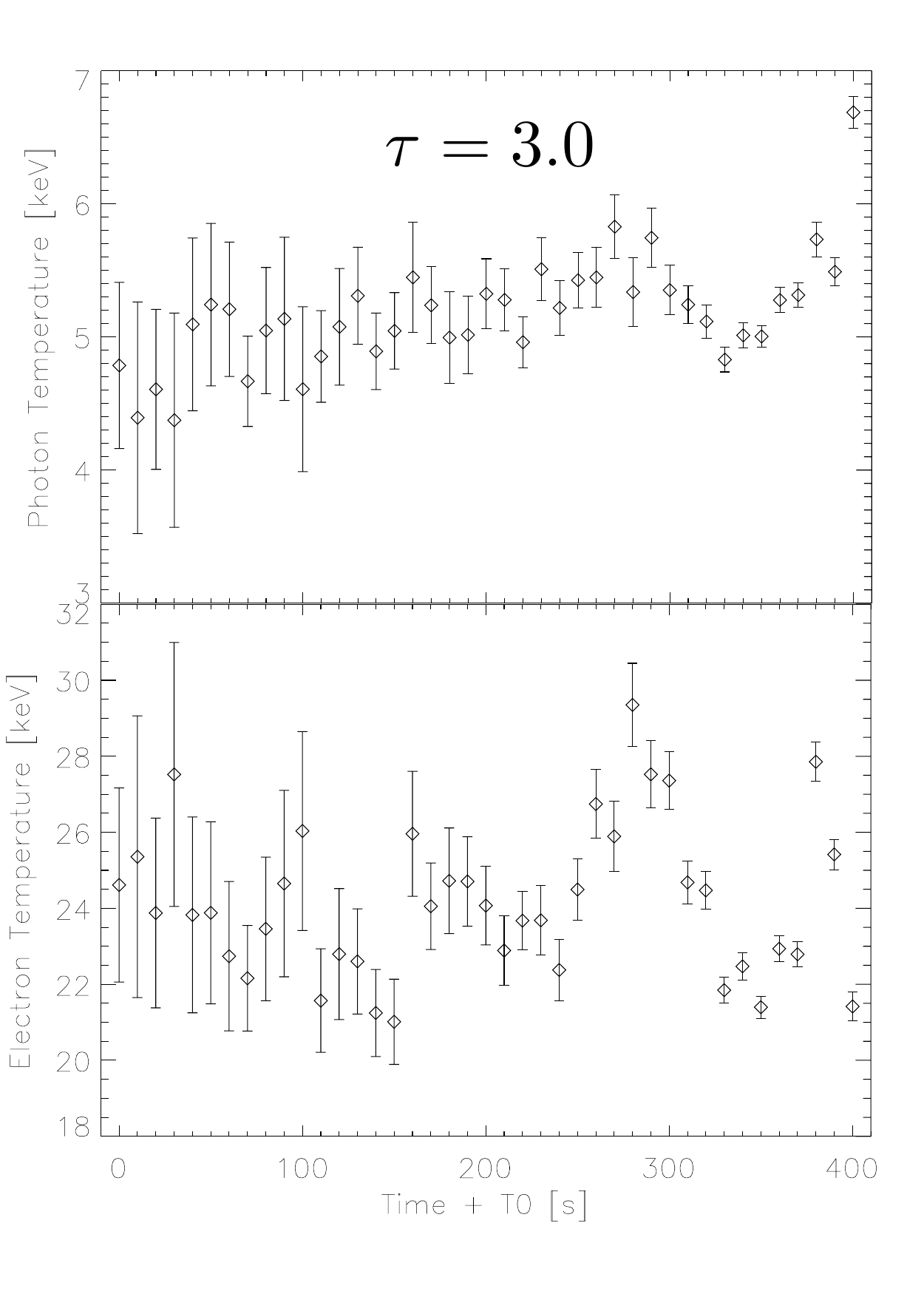}
\caption{The top figure shows the evolution of the photon seed temperature during the GBM triggered flare, bn150625400, while the bottom figure shows the evolution of the electron temperature.  The horizontal axis is time in seconds with T0 = 2015-06-25 09:33:43.6 UT.  The optical depth, $\tau$, is fixed to the best fit time integrated value.}
\label{fig:k0}
\end{centering}
\end{figure}

Time resolved spectral analysis was performed on a few bright triggered flares to examine spectral variation within a flare.
Ten second intervals were used for the time resolved analysis and an instrument response matrix was created for the centroid of each time interval.  Only the CompTT model was used to fit the time resolved data.  The trigger bn150625400 was chosen because it spanned a wide range of luminosities.  The triggers bn150626685 and bn150626751 were chosen for their high luminosity while bn150626156 was chosen for its moderate luminosity. For the individual triggers, the optical depth varied very little therefore the optical depth was frozen at the best parameter fit from the time integrated spectral analysis.  Figure \ref{fig:k0}  shows how the electron and photon temperature evolve during the bright flare,  bn150625400.  There is only a small segment of time between 280 and 350 s where the electron temperature and photon temperature appear correlated.  

%
\subsection{Temporal Data Analysis}

\subsubsection{Data Selection}
GBM CTTE NaI data between 8 and 100 keV were selected between 2015 June 15--27.  Times during SAA passage, times when V404 Cyg was occulted by the Earth  and times during high particle activity were excluded from the analysis.  The data selection resulted in good time intervals (GTIs) that were a few$\times$1000 seconds long.  On a 100 second cadence, the selected CTTE data were combined for all the detectors which had a source to detector bore sight angle of less than 60 degrees, and binned to 1 ms, to produce 100 second long light curves.  The light curves were Fourier transformed producing power spectra with a frequency range of 0.01--500 Hz.  The power spectra for each GTI were averaged to reduce the variance resulting in 214 power spectra. 

Although source and background count rates were not available for all observations, the power spectra that were studied in detail (see below), were all rms normalized \citep{Belloni_1990, Miyamoto_1991}. We note, however,  that the source and background rates used for the rms normalization are estimates based on the spectral analysis.

\subsubsection{Temporal Analysis Results}

\begin{figure}
\centerline{\includegraphics[width=0.9\columnwidth]{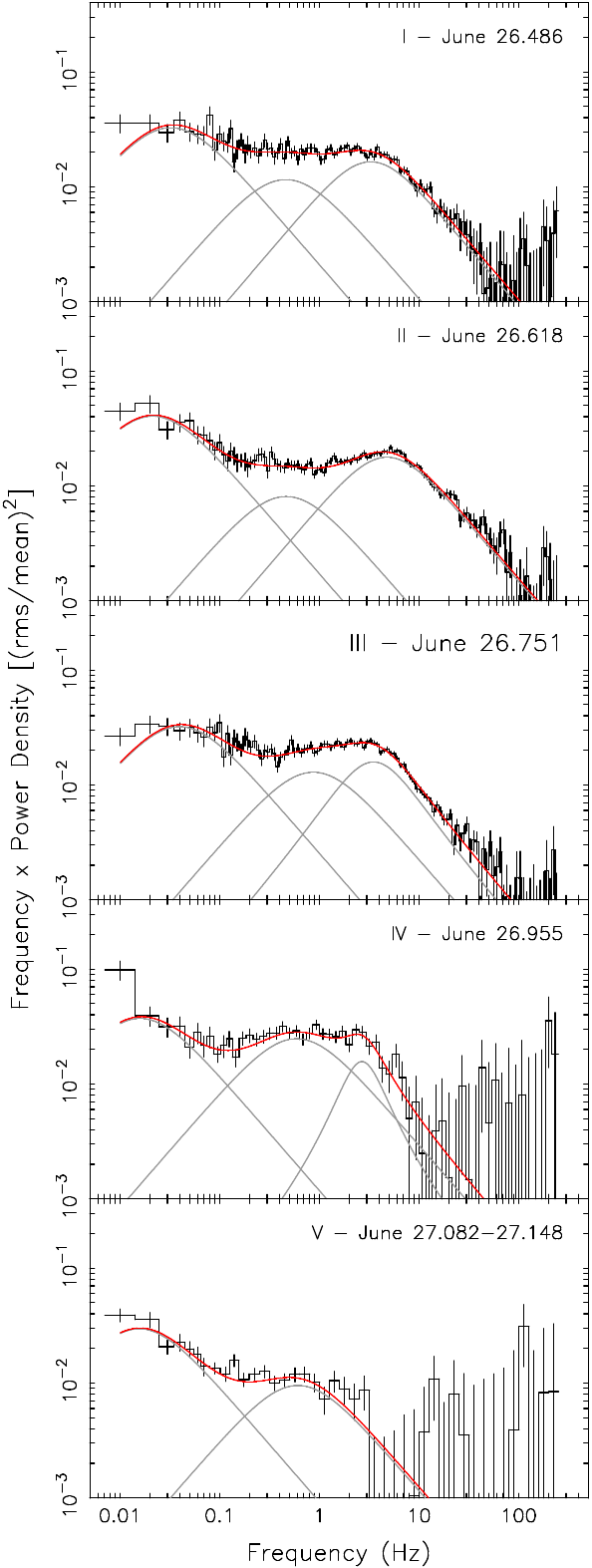}}
\caption{A sequence of  five power spectra of V404 Cyg from June 26--27. Models fits with two or three Lorentzian are shown in red; individual Lorentzians are plotted in gray. Corresponding dates are shown in the upper right corners. See Table \ref{tab:pds} for fit parameters.}\label{fig:pds}
\end{figure}

All 214 power spectra were visually inspected. For the ones from low-count-rate observations no significant power is seen. However, the power spectra from observations during bright flares are of high quality and show significant power. In Figure \ref{fig:pds} we show five normalized power spectra from the period of June 26--27, during which the source reached its peak brightness. They are representative of the power spectra during other flares. The power spectra are dominated by strong broad features, similar to  {\it Ginga} power spectra of V404 Cyg during its 1989 outburst (Oosterbroek et al. 1997). No narrow QPO features were seen. The integrated fractional rms in the 0.01--100 Hz band ranged from $\sim$35\% to $\sim$50\%. 

 Like the {\it Ginga} power spectra, we find that the Fermi/GBM power spectra of V404 Cyg can be fitted well with two or three broad Lorentzians, where we define the Lorentzians as ($P(\nu)=(r^2\Delta/\pi)[\Delta^2 +
(\nu-\nu_0)^2]^{-1}$). Here $\nu_0$ is the centroid
frequency, $\Delta$ the half-width-at-half-maximum, and $r$ the
integrated fractional rms (from $-\infty$ to $\infty$). Instead of
$\nu_0$ and $\Delta$ we will quote the frequency $\nu_{max}$ at which
the Lorentzian attains its maximum in $\nu P(\nu)$ and the quality
factor, $Q$, where $\nu_{max}=\nu_0 (1 + 1/4Q^2)^{1/2}$ and
$Q=\nu_0/2\Delta$. In some cases an additional power-law at low-frequencies provides a minor improvement to the fits, but for reasons of consistency this component was left out of our final model.

The Lorentzian fits to the power spectra are shown in Figure \ref{fig:pds}. As can be seen, the Lorentzians are well-separated in frequency. Small shifts in the frequencies of the Lorentzians are seen, but the overall shape of the power spectra remained the same, with perhaps an exception in the bottom panel of Figure \ref{fig:pds}. This was also mostly the case for the {\it Ginga} power spectra reported by Oosterbroek et al.\ (1997), who only observed one clear exception from the usual shape in their set of power spectra.  The $\nu_{max}$ ranges for the three Lorentzians are:  $\sim$0.016--0.04 Hz, $\sim$0.47--0.87 Hz, and $\sim$2.7--4.8 Hz. The $Q$-values of the broad noise features were less than 0.8 and in most cases were fixed at 0. The fractional rms amplitudes of the low-, mid-, and high-frequency Lorentzians were $\sim$31--36\%, $\sim$16--28\%, and $\sim$15--23\%, respectively. The full list of fit parameters is given in Table \ref{tab:pds}. Note that the quality of the June 27 power spectrum in Figure \ref{fig:pds} was not high enough to separately fit the two highest-frequency components, and only two Lorentzians were used to fit this power spectrum.  

\section{Discussion}
For a $10 M_\sun$ black hole the Schwarzschild radius is $R_S=3\times 10^{6}
$ cm and the Eddington luminosity is $L_{\rm EDD}=1.26\times 10^{39}
~{\rm erg~ s}^{-1}$.  The luminosities calculated for GBM data are from 10-1000 keV.  Significant flux is expected below 10 keV therefore our luminosities represent lower limits to the bolometric luminosity.  It is probable that V404 Cyg reached or exceeded the Eddington luminosity during the 2015 outburst.

\subsection{Spectral Modeling}
Initially a comptonized model  by \cite{Sunyaev_Titarchuk_1980} (CompST in XSPEC) was chosen to model the emission from hot coronal electrons up-scattering the cold accretion disk photons.  This model was chosen for its small number of parameters and its physical description of the emission region.  The model was often a poor description of the GBM data, especially during bright flares, and resulted in a large optical depth, tau (see Figure \ref{fig:CompTT}) which is inconsistent with the low electron temperatures.   Two absorption models, PHABS and PCFABS in XSPEC, were used to improve the model fit.  Both absorption models predicted extremely high absorption ($(100 - 1000)$ $ \times 10^{22} $ cm$^2$) that is unsupported by soft X-ray observations \citep{Motta_2015}.  A reflection model (REFLECT in XSPEC) was also used to attempt to improve the CompST fit.  The inclination of the the accretion disc was fixed to 56 degrees \citep{Khargharia_2010} and the redshift fixed to zero reflecting our proximity to the source.  The new model was again preferred over just the CompST model but resulted in an unphysical reflection component which exceeded the unreflected component which requires extreme relativistic light bending \citep{Fabian_2013}.  Other issues with the REFLECT*CompST model was, again, an excessively large optical depth that was often greater than 10.  The CompST model was abandoned in favor of the more flexible CompTT model.

\begin{figure}[h!]
\centerline{\includegraphics[width=0.9\columnwidth]{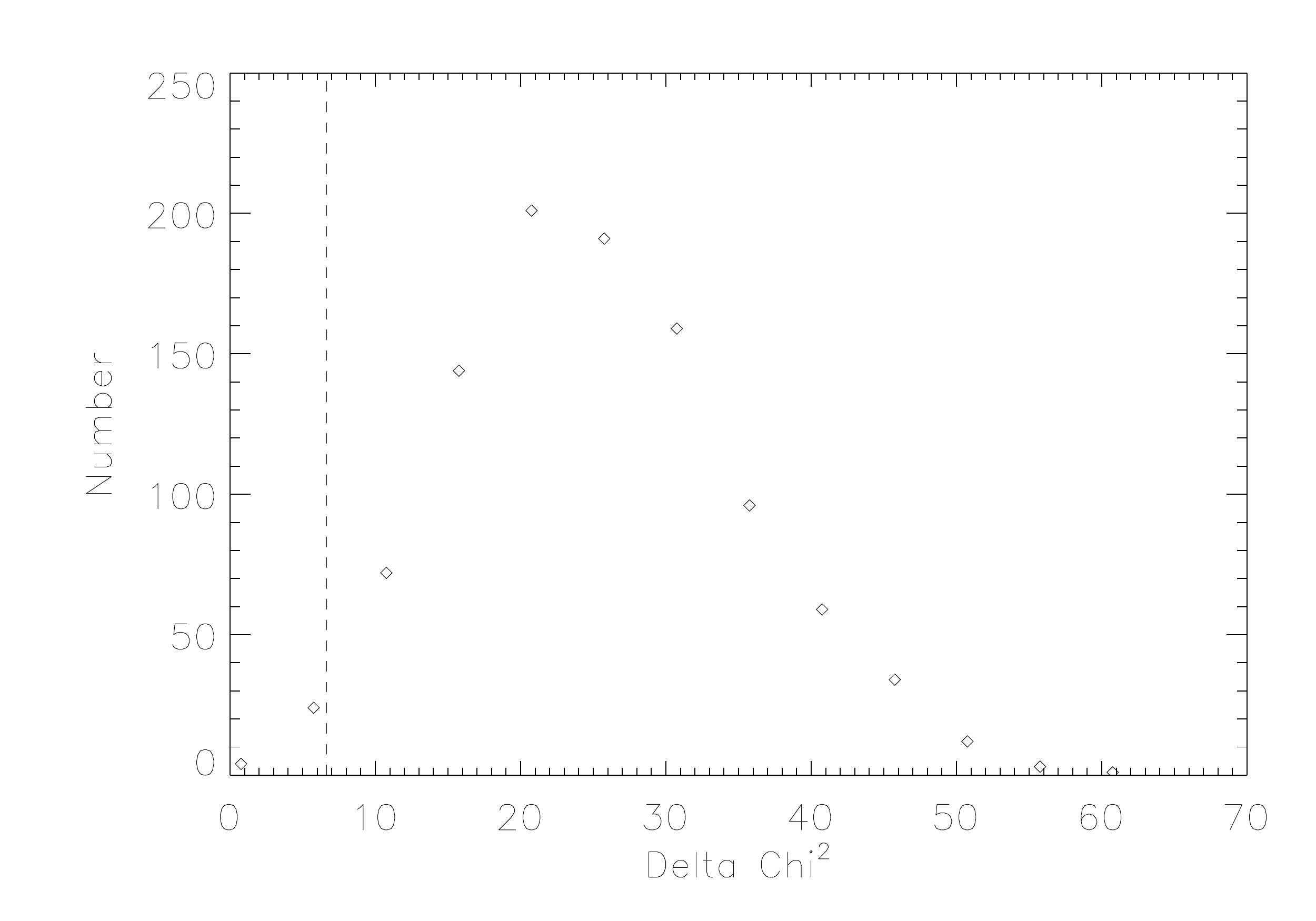}}
\caption{The distribution of the $\Delta\chi^{2}$ of the REFLECT*CompTT and CompTT model.  The dashed line is the 99\% rejection point.  The REFLECT*CompTT model can be rejected in all but the six simulated spectra.}
\label{fig:delta_chi}
\end{figure}

The CompTT model resulted in a  seed photon temperature that was too hot to originate from the accretion disk.  Including a reflection model and fixing the seed photon temperature to 1 keV resulted in reasonable fits for GBM data but the reflection component was often unphysically large ($\Omega/2\pi >> 1$) as well as the optical depth ($\overline{\tau} = 3.3$ with reflection as apposed to $\overline{\tau} = 1.6$ without reflection).  The spectrum from V404 Cyg has significant curvature around 20 keV which, assuming a comptonized model, may be fit with a high seed photon temperature or scattering a large number (on the order of the number of photons in the unreflected spectrum from 8 - 100 keV) of high energy photons down to lower energy and into the observers line of sight.  By fixing the reflection amplitude to 1.0 and fixing the seed photon temperature to 1.0 keV we refit all the spectra and compared the results to the CompTT fit with no reflection and the seed photon temperature as a free parameter.  Overall the CompTT model with the high seed photon temperature resulted in a better fit to the data.  Investigating further, we took the 12 seconds of the triggered event bn150626171 (see Figure  \ref{fig:CompTT}) and simulated 1000 sets of spectra (for detector n9 and na) using the best fit CompTT parameters.  We then simulated 1000 sets of spectra for the best fit parameters of the  REFLECT*CompTT model with the reflection amplitude and seed photon temperature fixed as before.  Comparing the $\Delta\chi^{2}$ of the resulting fits, all but a few of the fits using the REFLECT*CompTT model were rejected at the 99\% level and none resulted in a lower $\chi^{2}_{\nu}$ (see Figure \ref{fig:delta_chi}).  Although not statistically rigorous, the results are compelling enough that we can not reject the hot seed photons of the CompTT model. 

Relativistic light bending when the emission region is within a few gravitational radii of the black hole can produce reflection that exceeds $\Omega /2\pi = 1$ \citep{Fabian_2013}.  Fixing the reflection to $\Omega/2\pi = 1.5$ to take into account a large degree of relativistic light bending still results in a photon seed temperature that is consistently above 2 keV and averages 3.7 keV as opposed to 5.1 keV with no reflection.  Unfortunately we are unable to constrain both the reflection and the electron seed temperature.  Even though it is likely that there is a reflection component (Chandra observed broadened Fe K$\alpha$ lines, see \cite{King_2015}), GBM is not sensitive enough to constrain the reflection excess. The addition of the reflection component lowers the seed photon temperature for the REFLECT*CompTT fits by a marginal amount but does not alter the conclusion that these photons are too hot to be the thermal photons expected from an accretion disk.  Absorption might improve the spectral fits resulting in a thermal seed photon distribution but the absorption required is at least 100 times Galactic and neither XMM-Newton \citep{Rana_2015} nor Chandra \citep{King_2015} observed excessive absorption.  Absorption a few times Galactic will have little effect on our spectra above 8 keV so we omitted absorption in our spectral fits. 

\subsection{Physical Model for the Hard X-ray Emission}
\begin{figure}[!htbp]
\begin{center}

\includegraphics[width=0.9\columnwidth,angle=0]{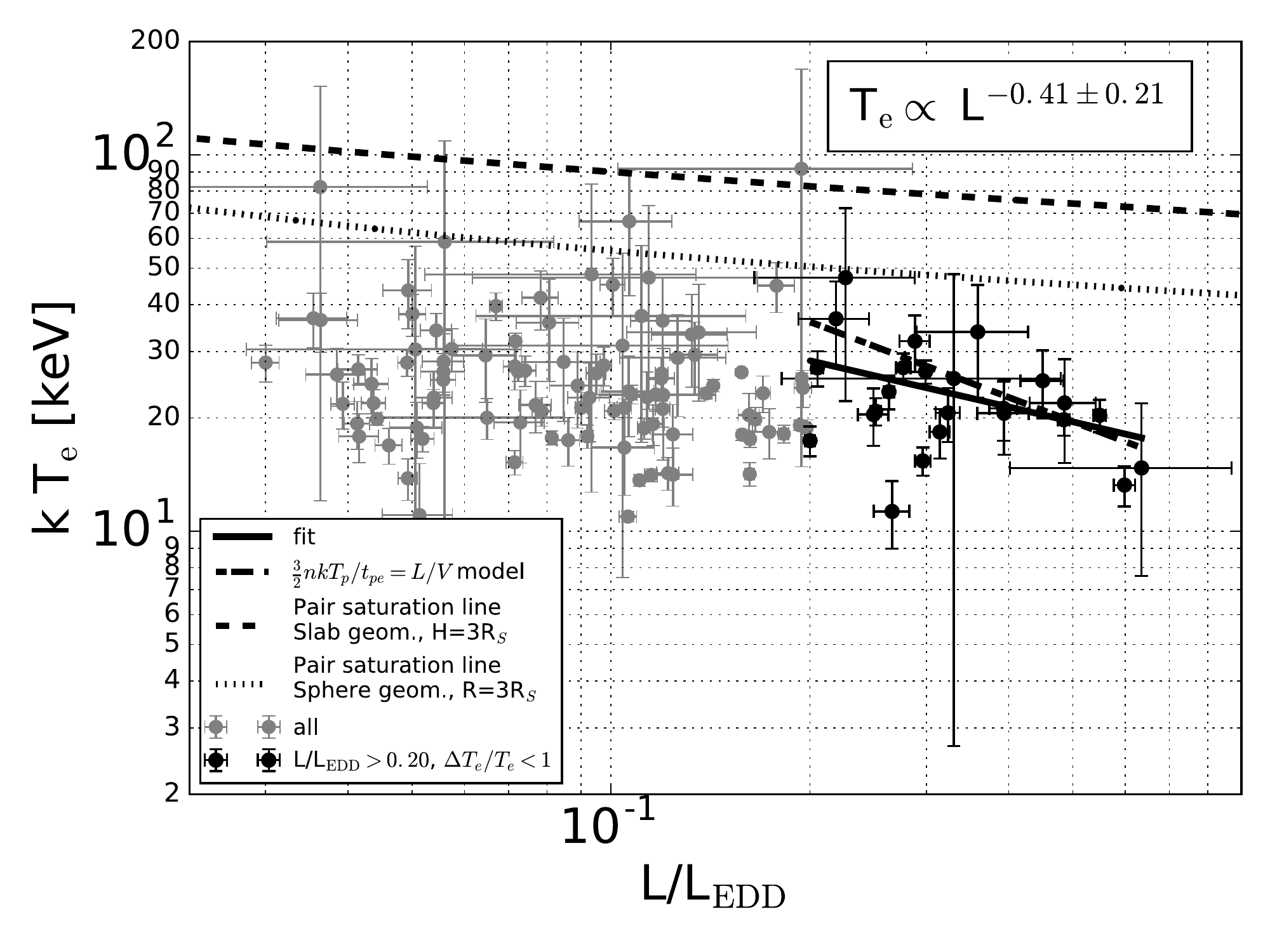}
\caption{Luminosity, $L/L_{\rm EDD}$, and electron temperature for all the triggered intervals.
Gray points mark all the intervals, black points indicate bright ($L>0.2L_{\rm
EDD}$) and reliable fits ($\chi^{2}_{\mu}<2$).}
\label{fig:int}
\end{center}
\end{figure}

\begin{figure}[!htbp]
\begin{center}
\includegraphics[width=0.9\columnwidth,angle=0]{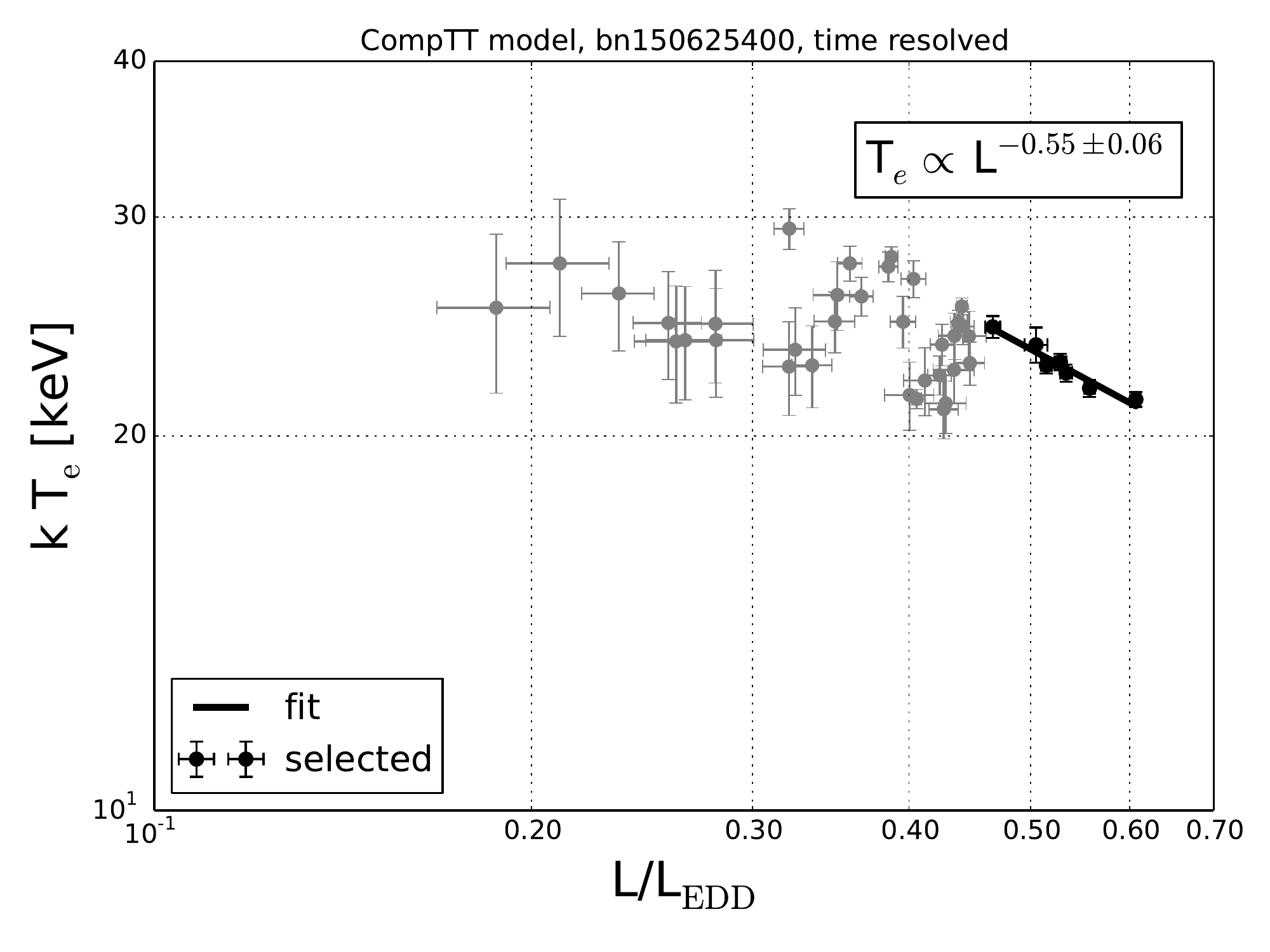}\\
\caption{Luminosity, $L/L_{\rm EDD}$, and electron temperature for an individual trigger.
Gray points mark all the intervals, black points indicate bright ($L>0.5L_{\rm
EDD}$) and reliable fits ($\chi^{2}{_\mu}<2$).}\label{fig:tres}
\end{center}
\end{figure}

\begin{figure}[!htbp]
\begin{center}
\includegraphics[width=0.9\columnwidth,angle=0]{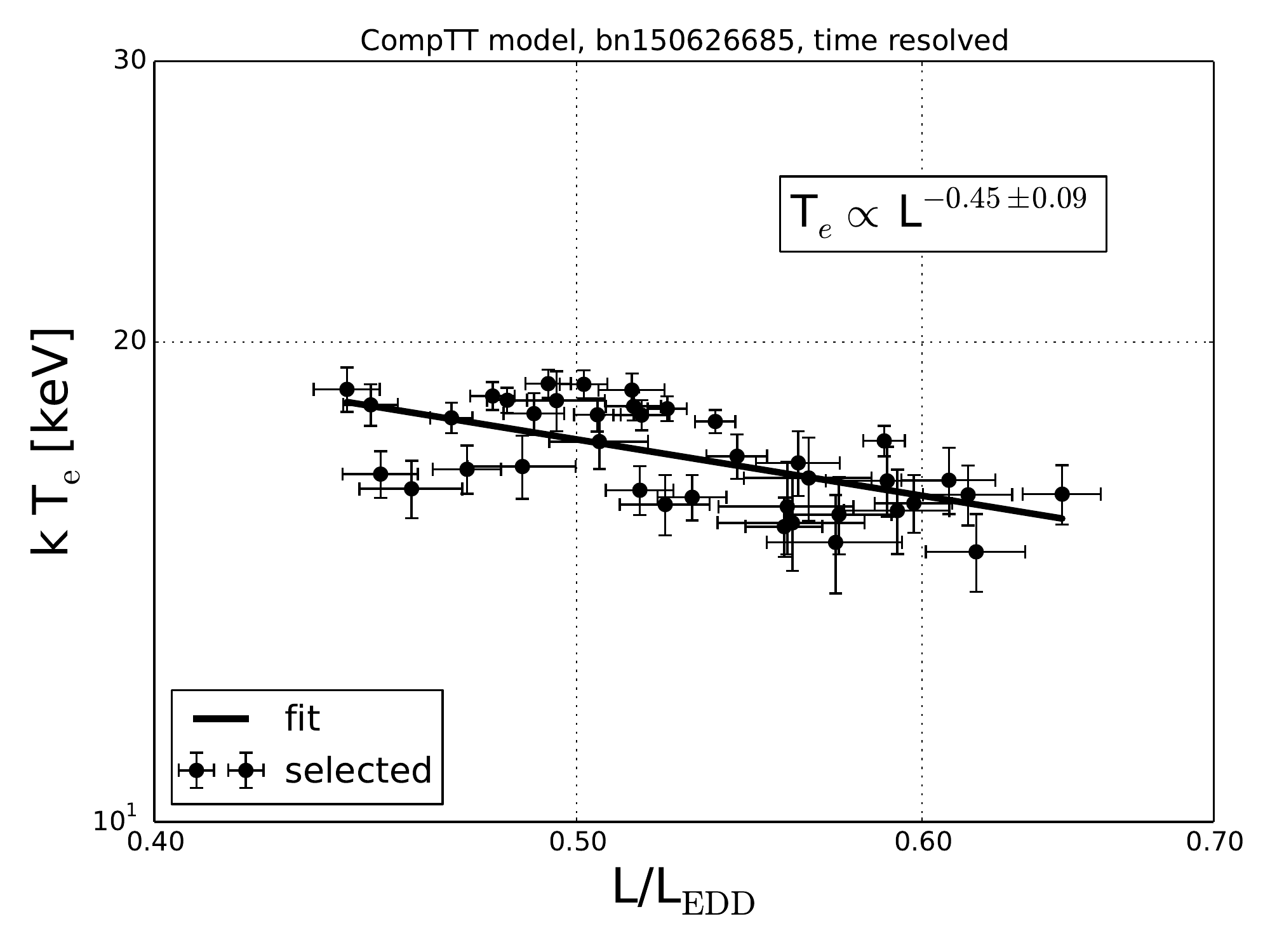}\\
\caption{A further example of a time resolved fit.  The anti-correlation exists for the entire flare because it remained above $0.45L_{Edd}$.}\label{fig:tres3}
\end{center}
\end{figure}

\begin{figure}[!htbp]
\begin{center}
\includegraphics[width=0.9\columnwidth,angle=0]{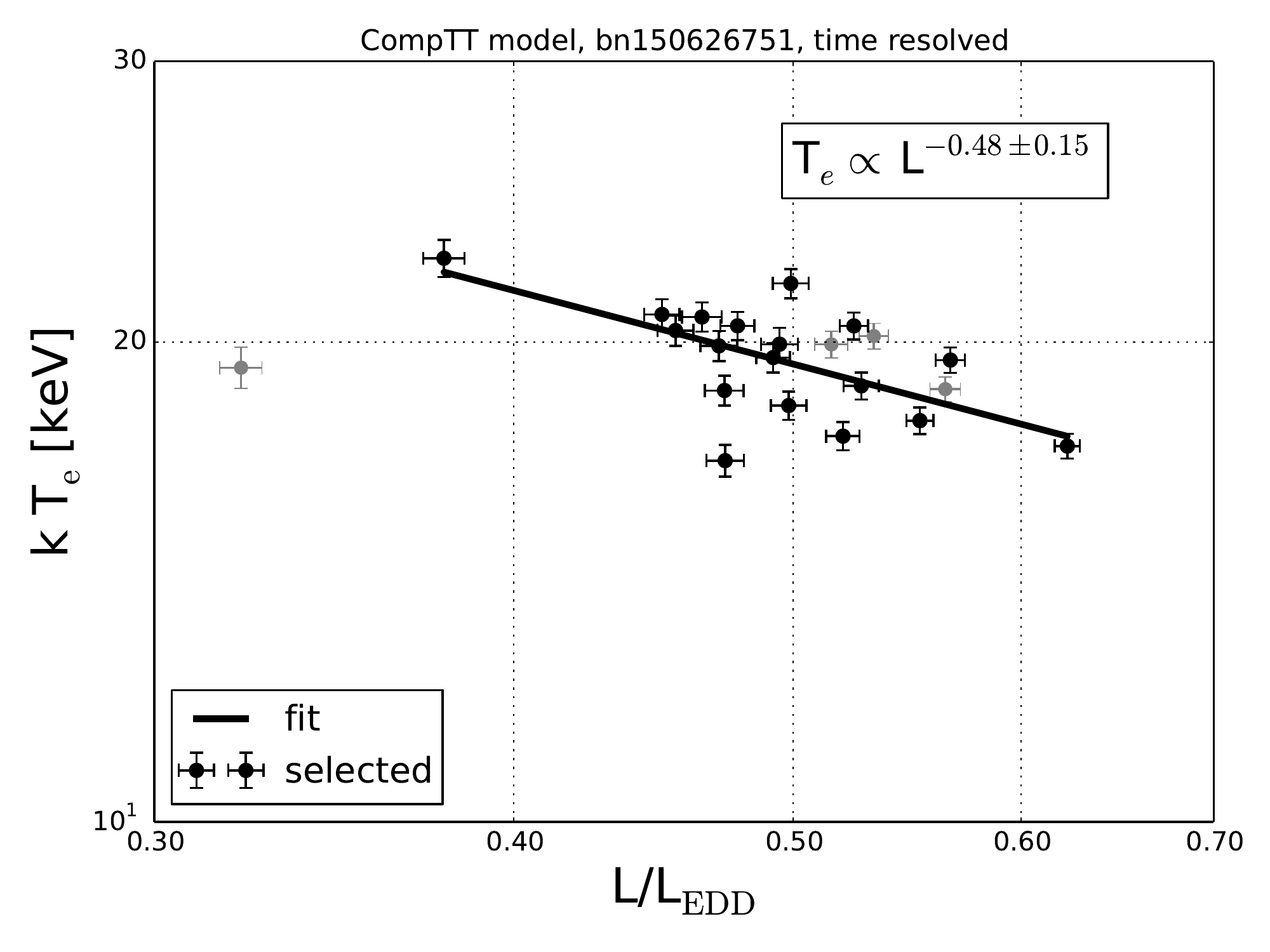}
\caption{Luminosity, $L/L_{\rm EDD}$, and electron temperature for individual triggers.
Gray points mark all the intervals, black points indicate bright ($L>0.5L_{\rm
EDD}$) and reliable fits ($\chi^{2}_{\mu}<2$).}\label{fig:tres4}
\end{center}
\end{figure}

We find a clear anti-correlation between the GBM flux and the electron
temperature of the CompTT model.  We present this behavior in the equivalent 
$L/L_{\rm EDD}-kT_e$ diagram.  This correlation is
present both in the brighter flares  (see Figures \ref{fig:tres},\ref{fig:tres3},\ref{fig:tres4})  and when considering the entire duration of
the current V404 Cyg flaring activity (see Figure \ref{fig:int}).  

Overall, the electron temperature values show large variations and no obvious
correlation  up to $L \lesssim 0.2 L_{\rm EDD}$. For $L >0.2 L_{\rm EDD}$ a
clear anti-correlation emerges, and the scatter in $T_e$ decreases visibly (see
Figure \ref{fig:int}).  Individual outbursts show a
similar behavior, when well sampled (trigger bn150625400 in Figure
\ref{fig:tres}).

The correlation between the electron temperature and luminosity has already
been noted in the case of GX 339-4 by \citet{Miyakawa+08lum}.  Using GBM
observations of V404 Cyg, we can populate a larger swath of the $L-kT_e$
diagram and find their interpretation valid here as well with minor
modifications.  We envision a population of protons in thermal equilibrium with
hot electrons which are in turn responsible for the inverse Compton
up-scattering of the soft thermal photons from the disk or base of the jet,
resulting in the gamma-ray photons.

In order to determine which processes drive the bursting activity, we calculate
relevant timescales.  For a case when protons have non-relativistic temperatures electron-proton relaxation timescale can be
calculated as \citep{Spitzer62,Dermer_1986}:
\begin{eqnarray}
t_{pe}&\approx& \frac{ \sqrt{2\pi} \left(\frac{kT_p}{m_p c^2} + \frac{kT_e}{m_e
c^2}\right)^{\frac{3}{2}} }{2 n_e \sigma_T c \ln \Lambda} \frac{m_p}{m_e}
\approx\\
 &\approx&1.3 \times 10^{-3} \left(\frac{R}{10~R_S}\right) \left(\frac{kT_e}{30~ {\rm
  keV}}\right)^{\frac{3}{2}} \left(\frac{\tau}{1.6}\right)^{-1} {\rm s},
\end{eqnarray}
where $n_e$ is the particle number density, $\ln \Lambda = 16.1 + \ln [(k
T_e/~30 {\rm keV}) (n_e/10^{17}{\rm cm}^{-3})^{-1/2}]$ is the Coulomb
logarithm, $T_e$ and $T_p$ are the electron and proton temperatures
respectively. Electrons and protons establish a Maxwellian distribution on this
timescale. For the numerical value, we have assumed $T_e>\frac{m_e}{m_p}T_p$
and $\tau\approx R n_e \sigma_T$. Henceforth, we use $10R_S$ for the size of
the emission region.

%

Electrons will lose energy to soft photons through inverse Compton scattering
on a timescale
\begin{eqnarray}
t_{\rm IC}&=&\frac{ 3 m_e c }{8 \sigma_T u_\gamma \tau} \\
	&=& 10^{-6} \left(\frac{L/L_{\rm EDD}}{0.2} \right)^{-1} \left(\frac{f}{9}
	\right)^{-1}  \left(\frac{R}{10~R_S}
\right)^{2} \left(\frac{\tau}{1.6}\right)^{-1} {\rm s}
\end{eqnarray}
where $u_\gamma=L f/4\pi R^2 c$ is the energy density of soft photons entering
the corona, $f$ is the ratio between the luminosity of the soft, disk component
($L_s$) and the upscattered hard X-ray luminosity, observed by GBM ($L$). $f$
follows e.g. from the derivation of \cite{Pietrini_1995}, and yields
$f=L_s/L\approx <(10 \theta_e \tau)^{-4}>\approx9$, and approximately constant for
luminosities above $0.2 L_E$  where $\theta=kT_e/m_e c^2$.

The advection timescale, or the time in which electrons are swallowed by the
black hole \citep{Mahadevan+97adaf} can be calculated e.g. in the advection
dominated accretion flow model \citep{Narayan+94adaf}:
\begin{eqnarray}
t_{\rm adv} &\approx& \frac{R^2}{\alpha H^2 \sqrt{G M/R^3}}\\
	&=& 0.4 \left(\frac{R/H}{0.2}\right)^2 \left(\frac{R}{10~R_S}\right)^{3/2}
	\left(\frac{\alpha}{0.3}\right)^{-1} {\rm s},
\end{eqnarray}
where $H$ is the height of the disk, assumed to be a fraction 0.2 of the radius,
and $\alpha$ is the viscosity parameter scaled to 0.3.

Out of these three timescales, the Compton cooling is the shortest. Based on
the above equations the proton-electron equilibrium timescale is shorter than
the advection timescale indicating the former is the more efficient process.
However the value of $H$, $\alpha$ and the emission region size assumed here,
$10 R_S$ have large uncertainties allowing $t_{\rm adv}$ to be of the same
order as $t_{\rm pe}$ (e.g. for $H/R\approx\alpha \approx1$, and $R \approx 3
R_S$).  If this is the case, about an equal fraction of proton energy will be
available {\bf to the electrons for} IC up scattering and for advection. As the luminosity increases,
IC cooling becomes more effective, decreasing the temperature of the electrons.
Colder electrons result in more effective proton-electron collisional
relaxation, while the advection efficiency does not vary with electron
temperature. Thus eventually the two main processes will be the electron proton
interaction and the IC cooling of the electrons.

If we assume a steady state, the energy transferred per unit volume and unit
time from protons to electrons will equal the energy lost by electrons through
Compton cooling. Following \citet{Inoue94ginga,Miyakawa+08lum}, yields:
\begin{equation}
\label{eq:eq}
\frac{3/2 n_e k T_p}{t_{\rm pe}} = 4\theta_e u\gamma n_e \sigma_T c.
\end{equation}

From Equation \ref{eq:eq}, using the expression for $t_{\rm pe}$, we find
$T_e\propto L^{-2/5}$, which is in remarkable agreement with both the time
resolved spectra (see Figures \ref{fig:tres},\ref{fig:tres3},\ref{fig:tres4}) and  considering all the bursts (see
Figure \ref{fig:int}).

Here, we neglected variation of the proton temperature and optical depth.
Since the proton-electron heating timescale is shorter than the advection
timescale, in the absence of heating, the protons might cool down to lower
temperatures before being advected. To see if the proton temperature is
independent of other parameters (e.g. the luminosity) we calculate it following
\cite{Malzac_2008}  by defining the Coulomb compactness, neglecting pair
contributions:
\begin{equation}
l_C=\sqrt{8\pi} \tau^2 \ln \Lambda \frac{k (T_p-T_e)}{m_p c^2}  \left(\frac{k
T_p}{m_p c^2} +\frac{k T_e}{m_e c^2}\right)^{-3/2}.
\end{equation}
We fix $l_C=473$ by using the average $\tau=1.95$, $T_e=26$ keV and $T_p = 46 $
MeV (from $G M /R =k T_p/m_p$, $R=10R_S$), then solve for the proton
temperature in the $kT_p/m_p c^2<<1$ limit.  We get $T_p=T_e (1+8.6
\sqrt{7\theta_e} l_C/\tau^2)$. The average of the proton temperatures
is $<T_p>=15.2$ MeV, indeed somewhat lower than from the virial type energy considerations but higher than \cite{Droulans_2010}. 
The Pearson correlation coefficient between $\log L$ and $\log T_p$ is 0.14,
with p-value p=0.11, consistent with no correlation between L and $T_p$. This
method of determining $T_p$ is appropriate for an estimate and for checking
correlation with the luminosity. Unfortunately, it 
uses the same principle by which we
link $t_{\rm pe}$ to the electron temperature and hence renders equation
\ref{eq:eq} meaningless  (yields $T_p\propto T_e^{3/2}$) for the purpose of
further addressing the electron temperature - luminosity correlation.  


 GBM is not sensitive to photons below $\sim 8$ keV. Some
fraction of the bolometric luminosity is emitted at this range and it is
difficult for us to account for that (but see f parameter at the beginning of this section for an estimate). Pair annihilation represents only a few percent \citep{Siegert_2016} of the overall luminosity and is also not considered here.  We can simply equate the luminosity
observed by GBM ($L$) in a spherical volume of 10 R$_S$, assumed to be
dominated by Comptonized emission, to the collisional power of the protons
($\frac{3}{2}nkT_p/t_{\rm pe}$). This will yield
\begin{equation}
kT_e=36.2 {\rm keV} \left(\frac{L}{0.2 L_E}\right)^{-2/3}.
\label{eq:eq2}
\end{equation}
We derived the normalization assuming $T_p=16$ MeV, $n=10^{17}$ cm$^{-3}$.  The
power law index is steeper than the fitted value, but still consistent within
the errors. Moreover, the expression in equation \ref{eq:eq2} gives a good
description of the data, especially at high luminosities.

This observed relation between the electron temperature and luminosity suggests that
during an outburst the electrons and protons are in thermal equilibrium and the
electrons are cooled by the Compton scattering of thermal photons from the
disk. V404 Cyg is yet another example among accreting black hole systems (both
black hole binaries and AGN) where the collisional heating of electrons by
protons is a dominant process.

{\it Alternative models}. 
Alternatively the anticorrelation between the luminosity and electron
temperature can be investigated considering runaway pair production (e.g.
Fabian et al 2015 ). This occurs for high compactness and electron temperature
as additional power introduced to the corona does not go towards increasing the
temperature but to the creation of pairs.  Defining the compactness,
$l=\sigma_T L/m_e c^3 R=2\pi m_p/m_e (L/L_E) (R/R_S)^{-1}$, for a spherical
corona, we have a limit corresponding to $l\approx 10 \theta_e^{-5/2}
e^{1/\theta_e}$. We have also adopted the model of
Stern et al 1995 who calculated the pair line for a slab geometry. We assume
$3R_S$ for the height and the radius of the slab and sphere cases respective.
Larger sizes will be less constraining.  We have plotted the runaway pair
production lines for sphere and slab geometry on Figure \ref{fig:int}. We
conclude that our observations do not violate this limit for realistic
assumptions for the emission radius, but this effect does not seem to drive the
observed correlation.
Another possibility to consider is the truncated disk model \citep{Done_2007, Sobolewska_2011}. In this
model the disk is truncated at an inner radius, and hot, spherical corona fills
the space closer to the black hole.  With increasing luminosity, the disk
truncation radius decreases. This results in more effective cooling of the hot
corona by disk photons.  Similar to \cite{Sobolewska_2011}, we consider a
radiatively efficient and inefficient flow \citep{Sharma_2007}
 and link the hard-to-soft luminosity ratio (equivalent to the hard to
soft compactness ratio) to the hard luminosity. We further use the expression
of \cite{Pietrini_1995} for the linking the hard-to-soft luminosity ratio to the
temperature  ( $L/L_s \propto (\theta_e \tau)^4$, see previous section.) In
this model we find $L/L_s\propto L^{-1/4} (\propto L^{-3/2})$ in the
radiatively inefficient (efficient) flow case. For the temperature dependence,
assuming constant optical depth, we find $L\propto T_e^{-1/8}$ for the
inefficient flow and $L\propto T_e^{-3/8}$ for the efficient flow. The latter,
radiatively efficient flow case is consistent with our observed correlation at
high luminosities suggesting the truncated disk model is viable in explaining
the observations.


Similarly to \citet{Roques_2015}, the seed photon temperature obtained
from the CompTT model is unusually high.  The highest temperature from a
thermally radiating  disk is $kT_{\rm photon}\lesssim 1$ keV, while we have an
approximately constant $kT_{\rm photon} = 5.9\pm1.3$ keV.  This suggests the
seed photons might not originate solely from the disk, but from another
source as well (e.g. synchrotron photons from the jet \citep{Markoff+05syn}).    
A high seed photon temperature (7 keV) was also measured  using INTEGRAL data \citep{Roques_2015,Natalucci_2015}.

We note that for the time resolved spectra the onset of the correlation appears
to start at higher luminosities ($0.35-0.5 L_{\rm EDD}$, (see Figures
\ref{fig:tres}, \ref{fig:tres3}, \ref{fig:tres4}) while the time integrated correlation is valid for
$L>0.2 L_{\rm EDD}$. This can be explained by the longer integration times for
the data points ($\sim$500s as apposed to 10s), resulting in more
accurate spectral parameters for $L/L_{\rm EDD}$ in the range of 0.2-0.35. For
the time resolved cases the error on the parameters increases for this range
of luminosities, suppressing the correlation until sufficient photons are measured.  Restricting the $L/L_{\rm EDD} > 0.4$ substantially steepens the relationship for the time integrated results.

\subsection{Temporal Analysis}

\begin{figure}[!h]
\centerline{\includegraphics[width=0.9\columnwidth]{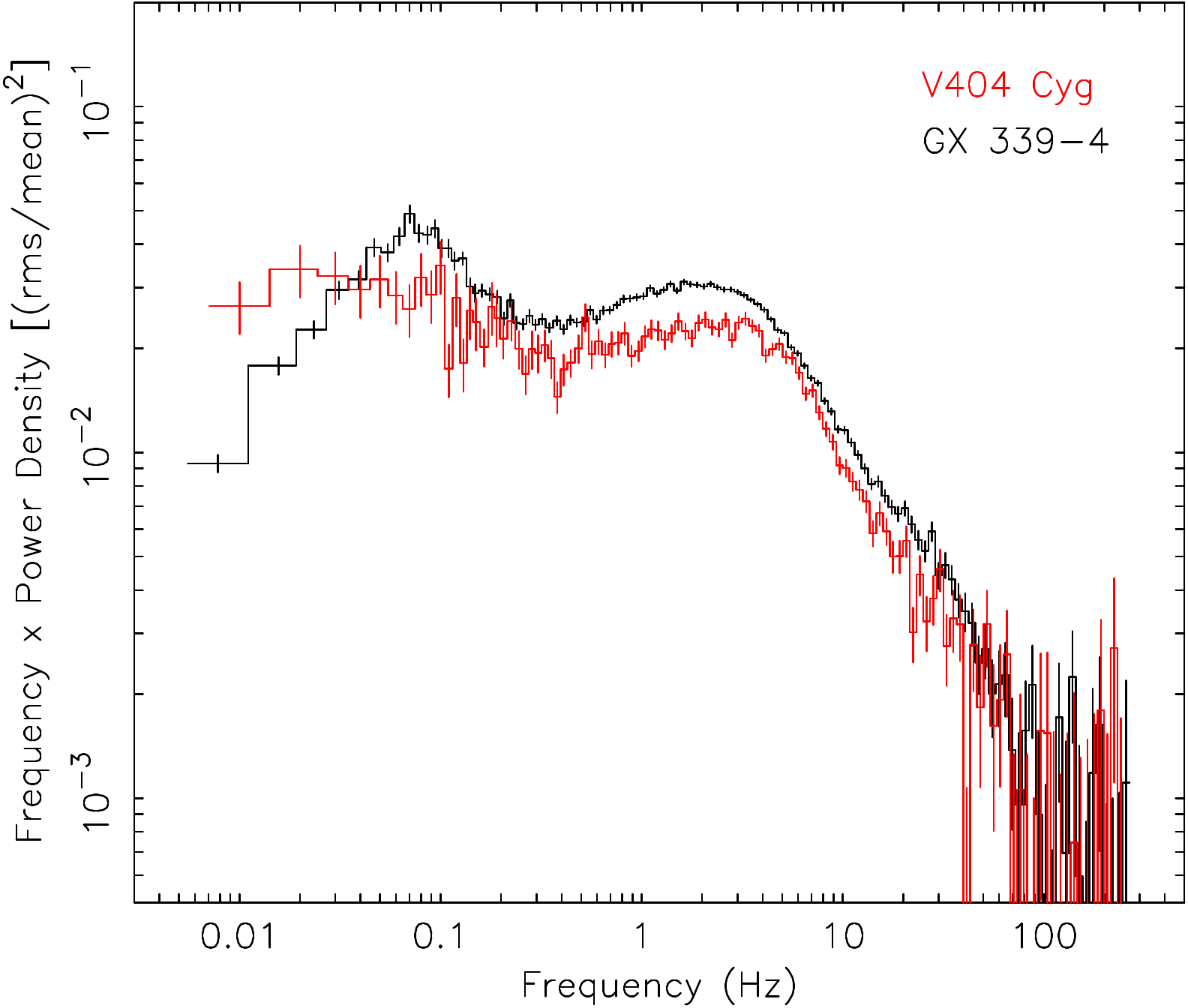}}
\caption{A comparison of an averaged {\it RXTE} power spectrum from the hard state of GX 339--4 (black) with a Fermi/GBM power spectrum of V404 Cyg from June 26.751 (red).}\label{fig:compare}
\end{figure}

\begin{center}
\begin{deluxetable*}{cccccccccccc}
\tabletypesize{\scriptsize}
\tablewidth{0pt}
\tablecaption{Power Spectral Fit Parameters\label{tab:pds}} 
\tablehead{
\colhead{} & 
\multicolumn{3}{c}{L$_{\rm low}$} & \colhead{} &
\multicolumn{3}{c}{L$_{\rm mid}$} & \colhead{} &
\multicolumn{3}{c}{L$_{\rm high}$}  \bigstrut[b] \\
\cline{2-4} \cline{6-8} \cline{10-12}
\colhead{} & 
\colhead{$\nu_{\rm max}$} & \colhead{$Q$} &  \colhead{rms} & \colhead{} &
\colhead{$\nu_{\rm max}$} & \colhead{$Q$} &  \colhead{rms} & \colhead{} &
\colhead{$\nu_{\rm max}$} & \colhead{$Q$} &  \colhead{rms} \bigstrut[t] \\
\colhead{PDS No.$^a$} & 
\colhead{(Hz)} & \colhead{} &  \colhead{(\%)} & \colhead{} &
\colhead{(Hz)} & \colhead{} &  \colhead{(\%)} & \colhead{} &
\colhead{(Hz)} & \colhead{} &  \colhead{(\%)}}

\startdata
I 	& (3.2$\pm$0.6)$\times10^{-2}$ & 0$^b$	& 32.1$\pm$1.5	& & 0.46$\pm$0.10	& 0$^b$	& 19.1$\pm$1.0	& & 3.3$\pm$0.3 	& (8$\pm$6)$\times10^{-2}$		& 22.0$\pm$1.1 \\
II	& (2.1$\pm$0.5)$\times10^{-2}$ & 0$^b$	& 36$\pm$3		& & 0.47$\pm$0.08	& 0$^b$	& 15.9$\pm$0.7	& & 4.78$\pm$0.16 & (7$\pm$3)$\times10^{-2}$		& 23.0$\pm$0.4 \\
III	& (4.0$\pm$0.5)$\times10^{-2}$ & 0$^b$	& 31.8$\pm$1.2	& & 0.87$\pm$0.15	& 0$^b$	& 15.9$\pm$0.7	& & 3.5$\pm$0.2 	& (2.8$\pm$0.5)$\times10^{-1}$	& 19.6$\pm$1.4 \\
IV	& (1.6$\pm$0.6)$\times10^{-2}$ & 0$^b$	& 34$\pm$5		& & 0.59$\pm$0.10	& 0$^b$	& 27.9$\pm$1.3	& & 2.7$\pm$0.3 	& (8$\pm$4)$\times10^{-1}$		& 15$\pm$3 \\
V	& (1.6$\pm$0.5)$\times10^{-2}$ & 0$^b$	& 31$\pm$4		& & 0.61$\pm$0.10	& 0$^b$	& 17.3$\pm$0.6	& & \nodata		& \nodata						& \nodata \\
\enddata 
\tablenotetext{a}{Number of power spectrum, as shown in Figure \ref{fig:pds}.}
\tablenotetext{b}{Value was fixed.}
\end{deluxetable*}
\end{center}

The Fermi/GBM power spectra of V404 Cyg show a strong similarity to those obtained with {\it Ginga} during the source's 1989 outburst. They are also similar in shape and strength to those of black hole X-ray binaries, such as Cyg X-1 and GX 339--4, in their hard states (Nowak 2000). The power spectra of these source are  dominated by broad Lorentzians as well, in the frequency range that we analyzed for V404 Cyg (0.01--500 Hz). In Figure \ref{fig:compare} we show a comparison between an averaged power spectrum from GX 339--4 in its hard state (from {\it RXTE} data) and one from V404 Cyg (June 26.71, see middle panel in Figure \ref{fig:pds}). As can bee seen, the shapes of these power spectra are very similar, especially at the high-frequency end. We suspect that relatively stronger variability at the low-frequency end in V404 Cyg may be the result of the strong flaring activity of V404 Cyg seen on the corresponding time scales. This additional power may hide the low-frequency break that is usually seen in the hard state (see, e.g., the GX 339--4 power spectrum in Figure \ref{fig:compare}). Overall, however, the shape and strength of the power spectra supports our earlier conclusion that V404 Cyg was observed in the hard state. 

\section{Conclusion}
There is no evidence in the spectral analysis for an onset of a state change during the observed time interval.  Spectral analysis indicates that the collisional heating of electrons by protons is the dominant process resulting in the observed Comptonized spectrum.  The seed photon temperature exceeds what is expected from an accretion disk and may be due to synchrotron photons from the base of the jet or perhaps there is some other method of energizing the photons of the inner disk.  In all, this outburst is very similar to the one that occurred in 1989 but remains remarkable among black hole outburst for its intrinsic high luminosity and high photon seed temperature.

\bibliographystyle{astron}

\bibliography{v404_Cyg_sub}

\begin{thebibliography}{}

\bibitem[\protect\astroncite{Agostinelli and et~al.}{2003}]{GEANT4}
Agostinelli, S. and et~al.: 2003,
\newblock {\em Nuclear Instruments and Methods in Physics Research Section A:
  Accelerators, Spectrometers, Detectors and Associated Equipment} {\bf
  506(3)}, 250

\bibitem[\protect\astroncite{{Arnaud}}{1996}]{xspec_1996}
{Arnaud}, K.~A.: 1996,
\newblock in G.~H. {Jacoby} and J. {Barnes} (eds.), {\em Astronomical Data
  Analysis Software and Systems V}, Vol. 101 of {\em Astronomical Society of
  the Pacific Conference Series}, p.~17

\bibitem[\protect\astroncite{{Barthelmy} et~al.}{2015}]{Barthelmy_2015}
{Barthelmy}, S.~D., {Chester}, M.~M., {Malesani}, D., and {Page}, K.~L.: 2015,
\newblock {\em GRB Coordinates Network} {\bf 17949}, 1

\bibitem[\protect\astroncite{{Belloni} and {Hasinger}}{1990}]{Belloni_1990}
{Belloni}, T. and {Hasinger}, G.: 1990,
\newblock {\em \aap} {\bf 227}, L33

\bibitem[\protect\astroncite{{Casares} and
  {Charles}}{1994}]{CasaresandCharles_1994}
{Casares}, J. and {Charles}, P.~A.: 1994,
\newblock in S. {Holt} and C.~S. {Day} (eds.), {\em The Evolution of X-ray
  Binariese}, Vol. 308 of {\em American Institute of Physics Conference
  Series}, p. 107

\bibitem[\protect\astroncite{{Connaughton} et~al.}{2015}]{Connaughton_2015}
{Connaughton}, V., {Briggs}, M.~S., {Goldstein}, A., {Meegan}, C.~A.,
  {Paciesas}, W.~S., {Preece}, R.~D., {Wilson-Hodge}, C.~A., {Gibby}, M.~H.,
  {Greiner}, J., {Gruber}, D., {Jenke}, P., {Kippen}, R.~M., {Pelassa}, V.,
  {Xiong}, S., {Yu}, H.-F., {Bhat}, P.~N., {Burgess}, J.~M., {Byrne}, D.,
  {Fitzpatrick}, G., {Foley}, S., {Giles}, M.~M., {Guiriec}, S., {van der
  Horst}, A.~J., {von Kienlin}, A., {McBreen}, S., {McGlynn}, S., {Tierney},
  D., and {Zhang}, B.-B.: 2015,
\newblock {\em \apjs} {\bf 216}, 32

\bibitem[\protect\astroncite{{Dermer}}{1986}]{Dermer_1986}
{Dermer}, C.~D.: 1986,
\newblock {\em \apj} {\bf 307}, 47

\bibitem[\protect\astroncite{{Done} et~al.}{2007}]{Done_2007}
{Done}, C., {Gierli{\'n}ski}, M., and {Kubota}, A.: 2007,
\newblock {\em \aapr} {\bf 15}, 1

\bibitem[\protect\astroncite{{Droulans} et~al.}{2010}]{Droulans_2010}
{Droulans}, R., {Belmont}, R., {Malzac}, J., and {Jourdain}, E.: 2010,
\newblock {\em \apj} {\bf 717}, 1022

\bibitem[\protect\astroncite{{Fabian}}{2013}]{Fabian_2013}
{Fabian}, A.~C.: 2013,
\newblock in C.~M. {Zhang}, T. {Belloni}, M. {M{\'e}ndez}, and S.~N. {Zhang}
  (eds.), {\em Feeding Compact Objects: Accretion on All Scales}, Vol. 290 of
  {\em IAU Symposium}, pp 3--12

\bibitem[\protect\astroncite{{Fender} et~al.}{2004}]{Fender04}
{Fender}, R.~P., {Belloni}, T.~M., and {Gallo}, E.: 2004,
\newblock {\em \mnras} {\bf 355}, 1105

\bibitem[\protect\astroncite{{Inoue}}{1994}]{Inoue94ginga}
{Inoue}, H.: 1994,
\newblock in T. {Courvoisier} and A. {Blecha} (eds.), {\em Multi-Wavelength
  Continuum Emission of AGN}, Vol. 159 of {\em IAU Symposium}, pp 73--82

\bibitem[\protect\astroncite{{Khargharia} et~al.}{2010}]{Khargharia_2010}
{Khargharia}, J., {Froning}, C.~S., and {Robinson}, E.~L.: 2010,
\newblock {\em \apj} {\bf 716}, 1105

\bibitem[\protect\astroncite{{King} et~al.}{2015}]{King_2015}
{King}, A.~L., {Miller}, J.~M., {Raymond}, J., {Reynolds}, M.~T., and
  {Morningstar}, W.: 2015,
\newblock {\em \apjl} {\bf 813}, L37

\bibitem[\protect\astroncite{{Mahadevan} and
  {Quataert}}{1997}]{Mahadevan+97adaf}
{Mahadevan}, R. and {Quataert}, E.: 1997,
\newblock {\em \apj} {\bf 490}, 605

\bibitem[\protect\astroncite{{Makino} et~al.}{1989}]{Makino_1989}
{Makino}, F., {Wagner}, R.~M., {Starrfield}, S., {Buie}, M.~W., {Bond}, H.~E.,
  {Johnson}, J., {Harrison}, T., and {Gehrz}, R.~D.: 1989,
\newblock {\em \iaucirc} {\bf 4786}, 1

\bibitem[\protect\astroncite{{Malzac}}{2012}]{Malzac_2012}
{Malzac}, J.: 2012,
\newblock {\em International Journal of Modern Physics Conference Series} {\bf
  8}, 73

\bibitem[\protect\astroncite{{Malzac} and {Belmont}}{2008}]{Malzac_2008}
{Malzac}, J. and {Belmont}, R.: 2008,
\newblock in {\em Microquasars and Beyond}, p.~7

\bibitem[\protect\astroncite{{Markoff} et~al.}{2005}]{Markoff+05syn}
{Markoff}, S., {Nowak}, M.~A., and {Wilms}, J.: 2005,
\newblock {\em \apj} {\bf 635}, 1203

\bibitem[\protect\astroncite{{McClintock} and {Remillard}}{2006}]{McClintock06}
{McClintock}, J.~E. and {Remillard}, R.~A.: 2006,
\newblock {\em {Black hole binaries}}, pp 157--213

\bibitem[\protect\astroncite{{Meegan} et~al.}{2009}]{Meegan_2009}
{Meegan}, C., {Lichti}, G., {Bhat}, P.~N., {Bissaldi}, E., {Briggs}, M.~S.,
  {Connaughton}, V., {Diehl}, R., {Fishman}, G., {Greiner}, J., {Hoover},
  A.~S., {van der Horst}, A.~J., {von Kienlin}, A., {Kippen}, R.~M.,
  {Kouveliotou}, C., {McBreen}, S., {Paciesas}, W.~S., {Preece}, R., {Steinle},
  H., {Wallace}, M.~S., {Wilson}, R.~B., and {Wilson-Hodge}, C.: 2009,
\newblock {\em \apj} {\bf 702}, 791

\bibitem[\protect\astroncite{{Miller-Jones} et~al.}{2009}]{Miller-Jones_2009}
{Miller-Jones}, J.~C.~A., {Jonker}, P.~G., {Dhawan}, V., {Brisken}, W.,
  {Rupen}, M.~P., {Nelemans}, G., and {Gallo}, E.: 2009,
\newblock {\em \apjl} {\bf 706}, L230

\bibitem[\protect\astroncite{{Miyakawa} et~al.}{2008}]{Miyakawa+08lum}
{Miyakawa}, T., {Yamaoka}, K., {Homan}, J., {Saito}, K., {Dotani}, T.,
  {Yoshida}, A., and {Inoue}, H.: 2008,
\newblock {\em \pasj} {\bf 60}, 637

\bibitem[\protect\astroncite{{Miyamoto} et~al.}{1991}]{Miyamoto_1991}
{Miyamoto}, S., {Kimura}, K., {Kitamoto}, S., {Dotani}, T., and {Ebisawa}, K.:
  1991,
\newblock {\em \apj} {\bf 383}, 784

\bibitem[\protect\astroncite{{Motta} et~al.}{2015}]{Motta_2015}
{Motta}, S., {Beardmore}, A., {Oates}, S., {Sanna}, N.~P.~M.~K.~A., {Kuulkers},
  E., {Kajava}, J., and {Sanchez-Fernanedz}, C.: 2015,
\newblock {\em The Astronomer's Telegram} {\bf 7665}, 1

\bibitem[\protect\astroncite{{Narayan} and {Yi}}{1994}]{Narayan+94adaf}
{Narayan}, R. and {Yi}, I.: 1994,
\newblock {\em \apjl} {\bf 428}, L13

\bibitem[\protect\astroncite{{Natalucci} et~al.}{2015}]{Natalucci_2015}
{Natalucci}, L., {Fiocchi}, M., {Bazzano}, A., {Ubertini}, P., {Roques}, J.-P.,
  and {Jourdain}, E.: 2015,
\newblock {\em \apjl} {\bf 813}, L21

\bibitem[\protect\astroncite{{Pietrini} and {Krolik}}{1995}]{Pietrini_1995}
{Pietrini}, P. and {Krolik}, J.~H.: 1995,
\newblock {\em \apj} {\bf 447}, 526

\bibitem[\protect\astroncite{{Rana} et~al.}{2015}]{Rana_2015}
{Rana}, V., {Loh}, A., {Corbel}, S., {Tomsick}, J.~A., {Chakrabarty}, D.,
  {Walton}, D.~J., {Barret}, D., {Boggs}, S.~E., {Christensen}, F.~E., {Craig},
  W., {Fuerst}, F., {Gandhi}, P., {Grefenstette}, B.~W., {Hailey}, C.,
  {Harrison}, F.~A., {Madsen}, K.~K., {Rahoui}, F., {Stern}, D., {Tendulkar},
  S., and {Zhang}, W.~W.: 2015,
\newblock {\em ArXiv e-prints}

\bibitem[\protect\astroncite{{Remillard} and {McClintock}}{2006}]{Remillard06}
{Remillard}, R.~A. and {McClintock}, J.~E.: 2006,
\newblock {\em \araa} {\bf 44}, 49

\bibitem[\protect\astroncite{{Rodriguez} et~al.}{2015}]{Rodriguez15}
{Rodriguez}, J., {Cadolle Bel}, M., {Alfonso-Garz{\'o}n}, J., {Siegert}, T.,
  {Zhang}, X.-L., {Grinberg}, V., {Savchenko}, V., {Tomsick}, J.~A.,
  {Chenevez}, J., {Clavel}, M., {Corbel}, S., {Diehl}, R., {Domingo}, A.,
  {Gouiff{\`e}s}, C., {Greiner}, J., {Krause}, M.~G.~H., {Laurent}, P., {Loh},
  A., {Markoff}, S., {Mas-Hesse}, J.~M., {Miller-Jones}, J.~C.~A., {Russell},
  D.~M., and {Wilms}, J.: 2015,
\newblock {\em \aap} {\bf 581}, L9

\bibitem[\protect\astroncite{{Roques} et~al.}{2015}]{Roques_2015}
{Roques}, J.-P., {Jourdain}, E., {Bazzano}, A., {Fiocchi}, M., {Natalucci}, L.,
  and {Ubertini}, P.: 2015,
\newblock {\em \apjl} {\bf 813}, L22

\bibitem[\protect\astroncite{{Sharma} et~al.}{2007}]{Sharma_2007}
{Sharma}, P., {Quataert}, E., {Hammett}, G.~W., and {Stone}, J.~M.: 2007,
\newblock {\em \apj} {\bf 667}, 714

\bibitem[\protect\astroncite{{Siegert} et~al.}{2016}]{Siegert_2016}
{Siegert}, T., {Diehl}, R., {Greiner}, J., {Krause}, M.~G.~H., {Beloborodov},
  A.~M., {Bel}, M.~C., {Guglielmetti}, F., {Rodriguez}, J., {Strong}, A.~W.,
  and {Zhang}, X.: 2016,
\newblock {\em \nat} {\bf 531}, 341

\bibitem[\protect\astroncite{{Sobolewska} et~al.}{2011}]{Sobolewska_2011}
{Sobolewska}, M.~A., {Papadakis}, I.~E., {Done}, C., and {Malzac}, J.: 2011,
\newblock {\em \mnras} {\bf 417}, 280

\bibitem[\protect\astroncite{{Spitzer}}{1962}]{Spitzer62}
{Spitzer}, L.: 1962,
\newblock {\em {Physics of Fully Ionized Gases}}

\bibitem[\protect\astroncite{{Sunyaev} and
  {Titarchuk}}{1980}]{Sunyaev_Titarchuk_1980}
{Sunyaev}, R.~A. and {Titarchuk}, L.~G.: 1980,
\newblock {\em \aap} {\bf 86}, 121

\bibitem[\protect\astroncite{{Titarchuk}}{1994}]{Titarchuk_1994}
{Titarchuk}, L.: 1994,
\newblock {\em \apj} {\bf 434}, 570

\bibitem[\protect\astroncite{{Wilson-Hodge} et~al.}{2012}]{WilsonHodge2012}
{Wilson-Hodge}, C.~A., {Case}, G.~L., {Cherry}, M.~L., {Rodi}, J.,
  {Camero-Arranz}, A., {Jenke}, P., {Chaplin}, V., {Beklen}, E., {Finger}, M.,
  {Bhat}, N., {Briggs}, M.~S., {Connaughton}, V., {Greiner}, J., {Kippen},
  R.~M., {Meegan}, C.~A., {Paciesas}, W.~S., {Preece}, R., and {von Kienlin},
  A.: 2012,
\newblock {\em \apjs} {\bf 201}, 33

\end{thebibliography}
\IfFileExists{\jobname.bbl}{}
 {\typeout{}
  \typeout{******************************************}
  \typeout{** Please run "bibtex \jobname" to optain}
  \typeout{** the bibliography and then re-run LaTeX}
  \typeout{** twice to fix the references!}
  \typeout{******************************************}
  \typeout{}
 }
  \setlength{\tabcolsep}{2pt}   
\onecolumngrid

\begin{center}
 \begin{deluxetable*}{lccccccccccccc}  
 
 \tabletypesize{\tiny}
 \tablewidth{0pt}
\tablecolumns{13}
\tablecaption{V404 Cyg Spectral Modeling Results}  

\tablehead{
\colhead{} & \\
\colhead{bn}  &  \colhead{Dur}& \colhead{Reflection amp.$^{a}$}& \colhead{e$^{-1}$ Temp.$^{a}$} & \colhead{Seed $\gamma$ Temp.$^{a}$}& \colhead{Optical Depth $^{a}$} & \colhead{Luminosity$^{a,b}$} & \colhead{$\chi^{2}$/dof} & \colhead{e$^{-1}$ Temp.$^{c}$} & \colhead{Seed $\gamma$ Temp.$^{c}$}& \colhead{Optical Depth $^{c}$} & \colhead{Luminosity$^{b,c}$} &\colhead{ $\chi^{2}$/dof} \\
\colhead{}  & \colhead{[s]} & \colhead{} & \colhead{[keV]} & \colhead{[keV]} & \colhead{} & \colhead{$10^{37}[$erg cm$^{-2}$]} & \colhead{} & \colhead{[keV]} & \colhead{[keV]}& \colhead{} & \colhead{$10^{37}[$erg cm$^{-2}$]} &\colhead{} \\
                     &                   &                  &                 &                  &                  & \colhead{10-1000 keV}  &                  &                 &                 &                   & \colhead{10-1000 keV} &         \\
}
 \startdata
150615791 &       31 & fixed at $\Omega/2\pi = 1$ & uncons & fixed at 1.0 & uncons & $5.7 \pm 1.1$ & 1.12 & uncons & $7.36 \pm 0.91$ & uncons & $5.5 \pm 1.0$ & 1.12\\
150615798 &      149 & fixed at $\Omega/2\pi = 1$ & $24.8 \pm 3.4$ & fixed at 1.0 & $1.67 \pm 0.29$ & $4.80 \pm 0.20$ & 1.26 & $21.7 \pm 3.1$ & $4.54 \pm 0.30$ & $1.71 \pm 0.34$ & $4.77 \pm 0.24$ & 1.26\\
150616734 &       14 & fixed at $\Omega/2\pi = 1$ & $32.3 \pm 5.8$ & fixed at 1.0 & $2.73 \pm 0.66$ & $4.78 \pm 0.56$ & 1.04 & $26.0 \pm 4.5$ & $4.5 \pm 1.8$ & $3.01 \pm 0.79$ & $4.67 \pm 0.52$ & 1.04\\
150616855 &       20 & fixed at $\Omega/2\pi = 1$ & $47.6 \pm 12.$ & fixed at 1.0 & $2.44 \pm 0.63$ & $4.6 \pm 1.8$ & 1.06 & $36.3 \pm 6.4$ & $3.5 \pm 2.2$ & $2.68 \pm 0.59$ & $4.41 \pm 0.66$ & 1.05\\
150616921 &      581 & fixed at $\Omega/2\pi = 1$ & $32.5 \pm 3.6$ & fixed at 1.0 & $2.47 \pm 0.33$ & $3.70 \pm 0.21$ & 0.95 & $28.0 \pm 3.1$ & $5.27 \pm 0.80$ & $2.38 \pm 0.40$ & $3.64 \pm 0.14$ & 0.95\\
150618710 &      162 & fixed at $\Omega/2\pi = 1$ & $54.9 \pm 19.$ & fixed at 1.0 & $1.38 \pm 0.48$ & $4.4 \pm 1.7$ & 1.11 & $36.7 \pm 6.0$ & $2.4 \pm 1.1$ & $1.90 \pm 0.31$ & $4.30 \pm 0.39$ & 1.10\\
150618763 &       10 & fixed at $\Omega/2\pi = 1$ & $53.4 \pm 39.$ & fixed at 1.0 & $1.0 \pm 1.0$ & $11.3 \pm 4.8$ & 1.23 & $48.1 \pm 35.$ & $6.6 \pm 1.0$ & uncons & $11.3 \pm 5.0$ & 1.23\\
150618834 &      173 & fixed at $\Omega/2\pi = 1$ & $44.5 \pm 3.6$ & fixed at 1.0 & $1.80 \pm 0.16$ & $8.24 \pm 0.19$ & 1.19 & $39.6 \pm 3.3$ & $6.46 \pm 0.29$ & $1.59 \pm 0.16$ & $8.13 \pm 0.21$ & 1.21\\
150619165 &      126 & fixed at $\Omega/2\pi = 1$ & $24.6 \pm 1.8$ & fixed at 1.0 & $3.15 \pm 0.38$ & $5.36 \pm 0.18$ & 0.89 & $21.8 \pm 1.8$ & $5.62 \pm 0.92$ & $2.94 \pm 0.50$ & $5.31 \pm 0.15$ & 0.89\\
150619173 &      139 & fixed at $\Omega/2\pi = 1$ & $33.8 \pm 2.9$ & fixed at 1.0 & $2.46 \pm 0.24$ & $6.12 \pm 0.28$ & 1.16 & $28.0 \pm 2.2$ & $4.69 \pm 0.61$ & $2.54 \pm 0.28$ & $5.97 \pm 0.19$ & 1.16\\
150619224 &       24 & fixed at $\Omega/2\pi = 1$ & $45.6 \pm 23.$ & fixed at 1.0 & $1.6 \pm 1.3$ & $5.2 \pm 1.5$ & 1.21 & uncons & $10.8 \pm 1.8$ & uncons & $5.5 \pm 1.4$ & 1.21\\
150619242 &       77 & fixed at $\Omega/2\pi = 1$ & $19.5 \pm 2.3$ & fixed at 1.0 & $2.63 \pm 0.45$ & $21.1 \pm 1.0$ & 1.00 & $18.3 \pm 2.7$ & $5.25 \pm 0.71$ & $2.36 \pm 0.62$ & $21.0 \pm 1.0$ & 1.00\\
150619301 &      248 & fixed at $\Omega/2\pi = 1$ & $27.5 \pm 13.$ & fixed at 1.0 & $1.23 \pm 0.91$ & $6.1 \pm 1.9$ & 1.20 & $30.3 \pm 26.$ & $5.65 \pm 0.57$ & uncons & $6.1 \pm 2.5$ & 1.20\\
150619309 &      155 & fixed at $\Omega/2\pi = 1$ & $24.1 \pm 2.0$ & fixed at 1.0 & $2.22 \pm 0.27$ & $20.64 \pm 0.45$ & 1.35 & $23.2 \pm 2.5$ & $6.38 \pm 0.41$ & $1.83 \pm 0.34$ & $20.64 \pm 0.69$ & 1.36\\
150619316 &      218 & fixed at $\Omega/2\pi = 1$ & $25.2 \pm 2.1$ & fixed at 1.0 & $2.03 \pm 0.24$ & $26.39 \pm 0.69$ & 1.56 & $47.1 \pm 24.$ & $7.31 \pm 0.21$ & $0.55 \pm 0.50$ & $27.5 \pm 10.$ & 1.49\\
150619342 &      286 & fixed at $\Omega/2\pi = 1$ & $27.92 \pm 0.78$ & fixed at 1.0 & $2.362 \pm 0.083$ & $23.79 \pm 0.26$ & 1.46 & $25.45 \pm 0.82$ & $5.39 \pm 0.17$ & $2.13 \pm 0.10$ & $23.62 \pm 0.25$ & 1.50\\
150619352 &       36 & fixed at $\Omega/2\pi = 1$ & $25.4 \pm 4.2$ & fixed at 1.0 & $2.73 \pm 0.62$ & $10.8 \pm 1.2$ & 1.13 & $24.3 \pm 5.5$ & $6.0 \pm 1.5$ & $2.21 \pm 0.85$ & $10.8 \pm 1.1$ & 1.13\\
150619420 &       15 & fixed at $\Omega/2\pi = 1$ & $25.2 \pm 4.4$ & fixed at 1.0 & $2.95 \pm 0.67$ & $8.0 \pm 1.6$ & 1.04 & $20.0 \pm 2.4$ & uncons & $3.67 \pm 0.71$ & $7.8 \pm 2.2$ & 1.04\\
150619427 &       96 & fixed at $\Omega/2\pi = 1$ & $30.8 \pm 2.9$ & fixed at 1.0 & $2.37 \pm 0.28$ & $9.10 \pm 0.32$ & 1.09 & $26.7 \pm 2.4$ & $5.34 \pm 0.61$ & $2.31 \pm 0.32$ & $9.00 \pm 0.32$ & 1.08\\
150619554 &      103 & fixed at $\Omega/2\pi = 1$ & $17.0 \pm 2.9$ & fixed at 1.0 & $2.58 \pm 0.86$ & $12.72 \pm 0.96$ & 1.68 & $16.6 \pm 4.2$ & $6.6 \pm 1.1$ & $2.0 \pm 1.2$ & $12.7 \pm 1.4$ & 1.69\\
150619561 &      194 & fixed at $\Omega/2\pi = 1$ & $27.6 \pm 1.5$ & fixed at 1.0 & $1.62 \pm 0.12$ & $36.36 \pm 0.43$ & 1.60 & $26.5 \pm 1.8$ & $5.51 \pm 0.14$ & $1.37 \pm 0.15$ & $36.37 \pm 0.56$ & 1.61\\
150619570 &      109 & fixed at $\Omega/2\pi = 1$ & $43.5 \pm 14.$ & fixed at 1.0 & $1.47 \pm 0.57$ & $10.0 \pm 2.0$ & 0.95 & $35.8 \pm 10.$ & $5.38 \pm 0.76$ & $1.48 \pm 0.58$ & $9.7 \pm 1.4$ & 0.95\\
150619580 &      180 & fixed at $\Omega/2\pi = 1$ & $38.1 \pm 34.$ & fixed at 1.0 & uncons & $5.5 \pm 2.5$ & 1.22 & uncons & $11.0 \pm 1.6$ & uncons & $5.4 \pm 2.4$ & 1.30\\
150619774 &      516 & fixed at $\Omega/2\pi = 1$ & $31.63 \pm 0.72$ & fixed at 1.0 & $2.641 \pm 0.069$ & $19.61 \pm 0.19$ & 1.38 & $26.43 \pm 0.58$ & $4.75 \pm 0.19$ & $2.695 \pm 0.089$ & $19.15 \pm 0.20$ & 1.43\\
150620136 &      275 & fixed at $\Omega/2\pi = 1$ & $24.51 \pm 0.65$ & fixed at 1.0 & $3.25 \pm 0.16$ & $16.73 \pm 0.26$ & 1.33 & $23.22 \pm 0.75$ & $5.70 \pm 0.44$ & $2.78 \pm 0.18$ & $16.94 \pm 0.27$ & 1.25\\
150620154 &      204 & fixed at $\Omega/2\pi = 1$ & $13.38 \pm 0.49$ & fixed at 1.0 & $3.38 \pm 0.16$ & $12.9 \pm 3.2$ & 1.61 & $10.93 \pm 0.32$ & $1.93 \pm 0.68$ & $4.75 \pm 0.29$ & $12.88 \pm 0.30$ & 1.60\\
150620171 &      163 & fixed at $\Omega/2\pi = 1$ & $26.1 \pm 2.3$ & fixed at 1.0 & $1.77 \pm 0.24$ & $13.42 \pm 0.22$ & 1.66 & $47.1 \pm 26.$ & $7.51 \pm 0.19$ & uncons & $13.8 \pm 6.3$ & 1.62\\
150620414 &      244 & fixed at $\Omega/2\pi = 1$ & $21.5 \pm 1.2$ & fixed at 1.0 & $2.43 \pm 0.17$ & $22.41 \pm 0.43$ & 1.32 & $18.13 \pm 0.96$ & $3.95 \pm 0.29$ & $2.71 \pm 0.22$ & $22.19 \pm 0.35$ & 1.35\\
150620421 &      293 & fixed at $\Omega/2\pi = 1$ & $16.94 \pm 0.70$ & fixed at 1.0 & $2.82 \pm 0.14$ & $13.56 \pm 0.59$ & 1.38 & $13.65 \pm 0.48$ & $2.77 \pm 0.31$ & $3.68 \pm 0.22$ & $13.41 \pm 0.21$ & 1.37\\
150620429 &      139 & fixed at $\Omega/2\pi = 1$ & $20.8 \pm 3.0$ & fixed at 1.0 & $2.26 \pm 0.40$ & $5.09 \pm 0.40$ & 1.13 & $17.8 \pm 2.6$ & $3.88 \pm 0.60$ & $2.47 \pm 0.53$ & $5.05 \pm 0.29$ & 1.13\\
150620503 &       22 & fixed at $\Omega/2\pi = 1$ & $33.5 \pm 9.2$ & fixed at 1.0 & $0.96 \pm 0.39$ & $43.9 \pm 6.9$ & 1.53 & $33.8 \pm 11.$ & $5.37 \pm 0.21$ & $0.72 \pm 0.46$ & $43.6 \pm 8.4$ & 1.52\\
150620533 &      203 & fixed at $\Omega/2\pi = 1$ & $22.2 \pm 1.5$ & fixed at 1.0 & $2.01 \pm 0.18$ & $23.53 \pm 0.42$ & 1.46 & $24.0 \pm 2.7$ & $5.73 \pm 0.21$ & $1.42 \pm 0.25$ & $23.72 \pm 0.62$ & 1.45\\
150620545 &       86 & fixed at $\Omega/2\pi = 1$ & $17.5 \pm 1.4$ & fixed at 1.0 & $2.01 \pm 0.22$ & $36.13 \pm 0.90$ & 1.30 & $15.3 \pm 1.3$ & $4.19 \pm 0.20$ & $2.15 \pm 0.29$ & $36.02 \pm 0.87$ & 1.30\\
150620552 &      535 & fixed at $\Omega/2\pi = 1$ & $23.08 \pm 0.97$ & fixed at 1.0 & $2.055 \pm 0.098$ & $19.46 \pm 0.32$ & 1.50 & $18.04 \pm 0.60$ & $3.20 \pm 0.17$ & $2.60 \pm 0.12$ & $19.19 \pm 0.27$ & 1.53\\
150620559 &      293 & fixed at $\Omega/2\pi = 1$ & $17.06 \pm 0.70$ & fixed at 1.0 & $3.00 \pm 0.15$ & $14.0 \pm 4.5$ & 1.50 & $14.10 \pm 0.55$ & $2.91 \pm 0.39$ & $3.77 \pm 0.27$ & $13.96 \pm 0.21$ & 1.50\\
150620567 &      435 & fixed at $\Omega/2\pi = 1$ & $18.0 \pm 1.3$ & fixed at 1.0 & $3.14 \pm 0.42$ & $4.97 \pm 0.13$ & 1.22 & $19.3 \pm 2.8$ & $6.80 \pm 0.65$ & $2.02 \pm 0.58$ & $5.01 \pm 0.20$ & 1.22\\
150620625 &      374 & fixed at $\Omega/2\pi = 1$ & $20.5 \pm 1.9$ & fixed at 1.0 & $1.91 \pm 0.26$ & $9.29 \pm 0.20$ & 1.65 & $21.6 \pm 3.3$ & $5.57 \pm 0.24$ & $1.40 \pm 0.36$ & $9.34 \pm 0.47$ & 1.63\\
150621282 &      101 & fixed at $\Omega/2\pi = 1$ & $11.9 \pm 3.1$ & fixed at 1.0 & $3.1 \pm 1.5$ & $6.22 \pm 0.79$ & 1.22 & $11.0 \pm 4.0$ & $5.7 \pm 2.5$ & uncons & $6.23 \pm 0.91$ & 1.23\\
150621289 &      440 & fixed at $\Omega/2\pi = 1$ & $18.4 \pm 2.1$ & fixed at 1.0 & $1.78 \pm 0.30$ & $10.46 \pm 0.28$ & 1.53 & $17.4 \pm 2.5$ & $4.95 \pm 0.23$ & $1.61 \pm 0.39$ & $10.46 \pm 0.54$ & 1.54\\
150621298 &      544 & fixed at $\Omega/2\pi = 1$ & $22.5 \pm 2.5$ & fixed at 1.0 & $1.83 \pm 0.29$ & $54.3 \pm 2.0$ & 1.45 & $25.0 \pm 5.1$ & $6.50 \pm 0.23$ & $1.16 \pm 0.40$ & $54.7 \pm 3.6$ & 1.49\\
150621327 &       65 & fixed at $\Omega/2\pi = 1$ & $21.3 \pm 1.3$ & fixed at 1.0 & $1.75 \pm 0.15$ & $30.51 \pm 0.34$ & 1.55 & $20.8 \pm 1.7$ & $5.22 \pm 0.14$ & $1.48 \pm 0.19$ & $30.56 \pm 0.43$ & 1.54\\
150621335 &       75 & fixed at $\Omega/2\pi = 1$ & $25.9 \pm 5.8$ & fixed at 1.0 & $1.66 \pm 0.50$ & $12.9 \pm 1.1$ & 1.09 & $23.5 \pm 5.6$ & $4.79 \pm 0.59$ & $1.59 \pm 0.59$ & $12.9 \pm 1.1$ & 1.09\\
150621344 &       94 & fixed at $\Omega/2\pi = 1$ & $22.8 \pm 1.5$ & fixed at 1.0 & $1.56 \pm 0.12$ & $19.96 \pm 0.38$ & 1.53 & $17.57 \pm 0.89$ & $3.19 \pm 0.12$ & $2.06 \pm 0.14$ & $19.72 \pm 0.24$ & 1.58\\
150621366 &      384 & fixed at $\Omega/2\pi = 1$ & $43.5 \pm 26.$ & fixed at 1.0 & uncons & $7.3 \pm 3.4$ & 1.58 & uncons & $7.16 \pm 0.24$ & uncons & $7.40 \pm 0.80$ & 1.65\\
150621657 &       86 & fixed at $\Omega/2\pi = 1$ & uncons & fixed at 1.0 & uncons & $42.2 \pm 15.$ & 1.05 & uncons & $5.38 \pm 0.35$ & uncons & $41.6 \pm 15.$ & 1.05\\
150621744 &      109 & fixed at $\Omega/2\pi = 1$ & $24.01 \pm 0.87$ & fixed at 1.0 & $2.47 \pm 0.10$ & $23.93 \pm 0.52$ & 1.51 & $19.12 \pm 0.58$ & $3.35 \pm 0.25$ & $2.99 \pm 0.14$ & $23.47 \pm 0.29$ & 1.53\\
150622002 &      319 & fixed at $\Omega/2\pi = 1$ & $44.5 \pm 14.$ & fixed at 1.0 & $0.85 \pm 0.36$ & $11.8 \pm 3.6$ & 1.22 & $26.2 \pm 3.2$ & $3.09 \pm 0.24$ & $1.55 \pm 0.24$ & $11.54 \pm 0.42$ & 1.23\\
150622018 &      159 & fixed at $\Omega/2\pi = 1$ & $24.7 \pm 2.4$ & fixed at 1.0 & $1.69 \pm 0.21$ & $11.02 \pm 0.30$ & 1.33 & $21.2 \pm 2.0$ & $4.55 \pm 0.22$ & $1.78 \pm 0.24$ & $10.95 \pm 0.25$ & 1.33\\
150622027 &      223 & fixed at $\Omega/2\pi = 1$ & $34.0 \pm 12.$ & fixed at 1.0 & $0.91 \pm 0.48$ & $10.3 \pm 2.7$ & 1.23 & $28.1 \pm 8.6$ & $4.30 \pm 0.24$ & $1.01 \pm 0.48$ & $10.3 \pm 1.1$ & 1.23\\
150622071 &       61 & fixed at $\Omega/2\pi = 1$ & $27.0 \pm 5.3$ & fixed at 1.0 & $2.10 \pm 0.52$ & $6.63 \pm 0.49$ & 1.35 & $22.6 \pm 3.8$ & $5.19 \pm 0.86$ & $2.23 \pm 0.59$ & $6.55 \pm 0.50$ & 1.35\\
150622079 &      299 & fixed at $\Omega/2\pi = 1$ & $19.8 \pm 2.1$ & fixed at 1.0 & $2.92 \pm 0.44$ & $5.67 \pm 0.31$ & 0.94 & $16.9 \pm 1.8$ & $4.40 \pm 0.85$ & $3.16 \pm 0.64$ & $5.60 \pm 0.28$ & 0.94\\
150622086 &      135 & fixed at $\Omega/2\pi = 1$ & $21.5 \pm 3.0$ & fixed at 1.0 & $3.37 \pm 0.95$ & $6.24 \pm 0.39$ & 1.19 & $18.8 \pm 3.1$ & $5.9 \pm 2.0$ & $3.2 \pm 1.3$ & $6.17 \pm 0.43$ & 1.19\\
150622187 &      417 & fixed at $\Omega/2\pi = 1$ & $18.3 \pm 1.5$ & fixed at 1.0 & $2.60 \pm 0.28$ & $8.75 \pm 0.37$ & 1.32 & $15.2 \pm 1.1$ & $3.64 \pm 0.50$ & $3.09 \pm 0.39$ & $8.68 \pm 0.21$ & 1.31\\
150622212 &      240 & fixed at $\Omega/2\pi = 1$ & uncons & fixed at 1.0 & uncons & $6.0 \pm 1.3$ & 1.29 & uncons & $5.11 \pm 0.20$ & uncons & $6.0 \pm 1.8$ & 1.32\\
150622225 &      179 & fixed at $\Omega/2\pi = 1$ & $33.0 \pm 5.4$ & fixed at 1.0 & $1.49 \pm 0.35$ & $14.44 \pm 0.78$ & 1.10 & $36.2 \pm 10.$ & $7.70 \pm 0.34$ & $0.94 \pm 0.46$ & $14.5 \pm 1.9$ & 1.16\\
150622256 &      270 & fixed at $\Omega/2\pi = 1$ & $26.1 \pm 8.0$ & fixed at 1.0 & $0.82 \pm 0.43$ & $12.6 \pm 1.3$ & 1.26 & $31.1 \pm 23.$ & $5.85 \pm 0.12$ & uncons & $12.6 \pm 5.4$ & 1.39\\
150622451 &      312 & fixed at $\Omega/2\pi = 1$ & $25.4 \pm 1.6$ & fixed at 1.0 & $2.32 \pm 0.18$ & $12.87 \pm 0.37$ & 1.13 & $21.2 \pm 1.2$ & $4.24 \pm 0.36$ & $2.53 \pm 0.22$ & $12.73 \pm 0.28$ & 1.12\\
150622470 &       86 & fixed at $\Omega/2\pi = 1$ & $22.6 \pm 4.6$ & fixed at 1.0 & $1.46 \pm 0.49$ & $14.5 \pm 1.1$ & 1.23 & $23.0 \pm 7.5$ & $6.23 \pm 0.36$ & $1.09 \pm 0.66$ & $14.5 \pm 2.2$ & 1.28\\
150622672 &      273 & fixed at $\Omega/2\pi = 1$ & $25.6 \pm 1.7$ & fixed at 1.0 & $1.52 \pm 0.15$ & $14.44 \pm 0.20$ & 1.65 & $26.2 \pm 2.6$ & $5.60 \pm 0.14$ & $1.17 \pm 0.19$ & $14.49 \pm 0.23$ & 1.65\\
150622684 &      208 & fixed at $\Omega/2\pi = 1$ & $19.6 \pm 1.2$ & fixed at 1.0 & $2.33 \pm 0.19$ & $11.22 \pm 0.21$ & 1.37 & $17.8 \pm 1.2$ & $4.67 \pm 0.27$ & $2.26 \pm 0.25$ & $11.18 \pm 0.18$ & 1.37\\
150623198 &       61 & fixed at $\Omega/2\pi = 1$ & $25.1 \pm 2.1$ & fixed at 1.0 & $1.92 \pm 0.20$ & $14.29 \pm 0.32$ & 1.35 & $23.0 \pm 2.2$ & $4.45 \pm 0.28$ & $1.82 \pm 0.25$ & $14.20 \pm 0.40$ & 1.36\\
150623272 &      461 & fixed at $\Omega/2\pi = 1$ & $15.2 \pm 1.5$ & fixed at 1.0 & $2.68 \pm 0.41$ & $5.99 \pm 0.19$ & 1.58 & $13.8 \pm 1.7$ & $5.10 \pm 0.49$ & $2.62 \pm 0.60$ & $5.98 \pm 0.20$ & 1.58\\
150623311 &      217 & fixed at $\Omega/2\pi = 1$ & $20.9 \pm 2.2$ & fixed at 1.0 & $2.13 \pm 0.31$ & $14.09 \pm 0.51$ & 1.37 & $19.2 \pm 2.3$ & $4.89 \pm 0.41$ & $2.02 \pm 0.40$ & $14.06 \pm 0.43$ & 1.37\\
150623339 &       53 & fixed at $\Omega/2\pi = 1$ & $51.3 \pm 9.4$ & fixed at 1.0 & $1.80 \pm 0.32$ & $6.4 \pm 1.3$ & 1.00 & $37.7 \pm 4.8$ & $4.51 \pm 0.76$ & $2.01 \pm 0.30$ & $6.08 \pm 0.33$ & 1.00\\
150623348 &      147 & fixed at $\Omega/2\pi = 1$ & $90.8 \pm 89.$ & fixed at 1.0 & uncons & $4.4 \pm 1.8$ & 1.06 & $82.0 \pm 70.$ & $4.9 \pm 1.5$ & uncons & $4.4 \pm 1.8$ & 1.06\\
150623443 &       65 & fixed at $\Omega/2\pi = 1$ & $23.6 \pm 4.0$ & fixed at 1.0 & $1.42 \pm 0.32$ & $47.8 \pm 2.8$ & 1.04 & $21.2 \pm 3.8$ & $4.32 \pm 0.24$ & $1.43 \pm 0.38$ & $47.7 \pm 3.1$ & 1.04\\
150623473 &      503 & fixed at $\Omega/2\pi = 1$ & uncons & fixed at 1.0 & uncons & $27.2 \pm 9.4$ & 1.66 & uncons & $7.03 \pm 0.25$ & uncons & $27.5 \pm 10.$ & 1.69\\
150623482 &       90 & fixed at $\Omega/2\pi = 1$ & $39.1 \pm 10.$ & fixed at 1.0 & $1.24 \pm 0.47$ & $18.7 \pm 2.2$ & 1.16 & uncons & $7.37 \pm 0.35$ & uncons & $19.1 \pm 8.7$ & 1.15\\
150623515 &      229 & fixed at $\Omega/2\pi = 1$ & $21.8 \pm 4.6$ & fixed at 1.0 & $1.48 \pm 0.43$ & $15.1 \pm 1.3$ & 1.26 & $18.0 \pm 3.6$ & $4.16 \pm 0.33$ & $1.69 \pm 0.49$ & $15.0 \pm 1.1$ & 1.27\\
150623526 &       44 & fixed at $\Omega/2\pi = 1$ & uncons & fixed at 1.0 & uncons & $14.70 \pm 0.22$ & 1.29 & uncons & $4.46 \pm 0.47$ & uncons & $14.76 \pm 0.28$ & 1.30\\
150623715 &      132 & fixed at $\Omega/2\pi = 1$ & $33.5 \pm 19.$ & fixed at 1.0 & $1.3 \pm 1.1$ & $14.0 \pm 6.2$ & 1.37 & uncons & $5.48 \pm 0.95$ & uncons & $14.5 \pm 3.2$ & 1.37\\
150623722 &       68 & fixed at $\Omega/2\pi = 1$ & $137.6 \pm 73.$ & fixed at 1.0 & $0.100 \pm 0.036$ & $23.29 \pm 0.66$ & 1.21 & $91.7 \pm 76.$ & $2.76 \pm 0.23$ & uncons & $23.6 \pm 11.$ & 1.22\\
150623741 &      208 & fixed at $\Omega/2\pi = 1$ & $54.7 \pm 47.$ & fixed at 1.0 & uncons & $11.9 \pm 5.4$ & 1.36 & uncons & $6.40 \pm 0.25$ & uncons & $11.8 \pm 5.3$ & 1.40\\
150623749 &      101 & fixed at $\Omega/2\pi = 1$ & $13.6 \pm 1.2$ & fixed at 1.0 & $2.30 \pm 0.38$ & $39.6 \pm 1.0$ & 1.66 & $25.4 \pm 22.$ & $6.24 \pm 0.12$ & uncons & $40.1 \pm 17.$ & 1.63\\
150624171 &       12 & fixed at $\Omega/2\pi = 1$ & $19.0 \pm 1.0$ & fixed at 1.0 & $2.23 \pm 0.19$ & $58.59 \pm 0.87$ & 1.54 & $19.7 \pm 1.8$ & $5.78 \pm 0.21$ & $1.66 \pm 0.26$ & $58.88 \pm 0.97$ & 1.54\\
150624178 &      125 & fixed at $\Omega/2\pi = 1$ & $21.4 \pm 2.9$ & fixed at 1.0 & $1.35 \pm 0.28$ & $30.3 \pm 1.1$ & 1.56 & $20.4 \pm 3.5$ & $4.79 \pm 0.17$ & $1.21 \pm 0.35$ & $30.3 \pm 1.5$ & 1.57\\
150624201 &      103 & fixed at $\Omega/2\pi = 1$ & $40.8 \pm 14.$ & fixed at 1.0 & $0.77 \pm 0.46$ & $18.7 \pm 5.4$ & 1.24 & uncons & $5.26 \pm 0.22$ & uncons & $19.2 \pm 7.6$ & 1.23\\
150624248 &      289 & fixed at $\Omega/2\pi = 1$ & $70.8 \pm 30.$ & fixed at 1.0 & $0.72 \pm 0.40$ & $12.6 \pm 4.2$ & 1.15 & $45.1 \pm 7.9$ & $4.26 \pm 0.20$ & $1.05 \pm 0.23$ & $12.23 \pm 0.53$ & 1.17\\
150624303 &       88 & fixed at $\Omega/2\pi = 1$ & $96.1 \pm 75.$ & fixed at 1.0 & $1.06 \pm 0.80$ & $7.4 \pm 3.5$ & 1.10 & uncons & $7.10 \pm 0.68$ & uncons & $7.3 \pm 3.3$ & 1.09\\
150624386 &      725 & fixed at $\Omega/2\pi = 1$ & $24.1 \pm 1.0$ & fixed at 1.0 & $2.68 \pm 0.14$ & $5.44 \pm 0.20$ & 1.72 & $19.86 \pm 0.76$ & $3.47 \pm 0.36$ & $3.10 \pm 0.18$ & $5.376 \pm 0.093$ & 1.70\\
150624445 &      266 & fixed at $\Omega/2\pi = 1$ & $40.8 \pm 14.$ & fixed at 1.0 & $1.19 \pm 0.60$ & $5.9 \pm 1.1$ & 1.29 & uncons & $7.47 \pm 0.49$ & uncons & $6.0 \pm 2.5$ & 1.28\\
150624568 &       77 & fixed at $\Omega/2\pi = 1$ & $48.5 \pm 22.$ & fixed at 1.0 & $0.77 \pm 0.60$ & $44.7 \pm 14.$ & 1.41 & uncons & $7.13 \pm 0.26$ & uncons & $45.8 \pm 14.$ & 1.40\\
150624585 &      137 & fixed at $\Omega/2\pi = 1$ & uncons & fixed at 1.0 & uncons & $4.6 \pm 1.9$ & 1.34 & uncons & $6.61 \pm 0.91$ & uncons & $4.60 \pm 0.51$ & 1.38\\
150624599 &      266 & fixed at $\Omega/2\pi = 1$ & $58.9 \pm 47.$ & fixed at 1.0 & uncons & $8.2 \pm 3.9$ & 1.11 & uncons & $5.69 \pm 0.33$ & uncons & $8.2 \pm 3.6$ & 1.11\\
150624724 &      118 & fixed at $\Omega/2\pi = 1$ & $23.0 \pm 2.6$ & fixed at 1.0 & $1.66 \pm 0.30$ & $15.15 \pm 0.38$ & 1.12 & $28.8 \pm 8.7$ & $6.92 \pm 0.24$ & $0.83 \pm 0.46$ & $15.3 \pm 2.2$ & 1.13\\
150625186 &      261 & fixed at $\Omega/2\pi = 1$ & $20.9 \pm 1.4$ & fixed at 1.0 & $3.51 \pm 0.38$ & $6.4 \pm 2.8$ & 1.15 & $17.5 \pm 1.3$ & $4.1 \pm 1.2$ & $3.84 \pm 0.72$ & $6.29 \pm 0.27$ & 1.16\\
150625263 &       36 & fixed at $\Omega/2\pi = 1$ & $17.4 \pm 1.1$ & fixed at 1.0 & $2.12 \pm 0.19$ & $24.26 \pm 0.40$ & 1.67 & $17.3 \pm 1.5$ & $5.05 \pm 0.17$ & $1.76 \pm 0.26$ & $24.31 \pm 0.56$ & 1.67\\
150625296 &      590 & fixed at $\Omega/2\pi = 1$ & $24.0 \pm 2.0$ & fixed at 1.0 & $1.75 \pm 0.19$ & $14.38 \pm 0.25$ & 1.70 & $25.4 \pm 3.3$ & $5.49 \pm 0.17$ & $1.29 \pm 0.26$ & $14.47 \pm 0.59$ & 1.67\\
150625363 &      635 & fixed at $\Omega/2\pi = 1$ & uncons & fixed at 1.0 & uncons & $10.6 \pm 4.8$ & 1.12 & uncons & $7.99 \pm 0.80$ & uncons & $10.92 \pm 0.12$ & 1.12\\
150625370 &      318 & fixed at $\Omega/2\pi = 1$ & $50.4 \pm 6.6$ & fixed at 1.0 & $1.41 \pm 0.22$ & $12.54 \pm 0.47$ & 1.63 & $66.5 \pm 24.$ & $8.09 \pm 0.29$ & $0.69 \pm 0.37$ & $12.9 \pm 2.6$ & 1.63\\
150625392 &       23 & fixed at $\Omega/2\pi = 1$ & $19.5 \pm 1.7$ & fixed at 1.0 & $2.31 \pm 0.29$ & $19.56 \pm 0.53$ & 1.28 & $20.3 \pm 2.8$ & $5.39 \pm 0.35$ & $1.74 \pm 0.40$ & $19.6 \pm 1.0$ & 1.27\\
150625400 &        8 & fixed at $\Omega/2\pi = 1$ & $21.7 \pm 4.1$ & fixed at 1.0 & $1.69 \pm 0.43$ & $47.8 \pm 3.0$ & 1.35 & $20.5 \pm 4.6$ & $4.76 \pm 0.36$ & $1.54 \pm 0.53$ & $47.8 \pm 3.9$ & 1.35\\
150625498 &      647 & fixed at $\Omega/2\pi = 1$ & $22.3 \pm 3.3$ & fixed at 1.0 & $2.34 \pm 0.53$ & $11.22 \pm 0.56$ & 1.28 & $22.6 \pm 5.3$ & $6.39 \pm 0.77$ & $1.76 \pm 0.69$ & $11.2 \pm 1.1$ & 1.28\\
150625530 &       60 & fixed at $\Omega/2\pi = 1$ & $17.7 \pm 1.9$ & fixed at 1.0 & $2.15 \pm 0.39$ & $8.80 \pm 0.26$ & 1.31 & $19.4 \pm 4.2$ & $6.36 \pm 0.35$ & $1.39 \pm 0.57$ & $8.85 \pm 0.88$ & 1.31\\
150625899 &       52 & fixed at $\Omega/2\pi = 1$ & $25.5 \pm 1.5$ & fixed at 1.0 & $3.09 \pm 0.24$ & $12.52 \pm 0.76$ & 1.19 & $20.8 \pm 1.0$ & $3.40 \pm 0.71$ & $3.58 \pm 0.33$ & $12.28 \pm 0.37$ & 1.17\\
150625921 &      437 & fixed at $\Omega/2\pi = 1$ & $47.1 \pm 44.$ & fixed at 1.0 & uncons & $8.7 \pm 4.0$ & 1.20 & uncons & $5.96 \pm 0.17$ & uncons & $8.7 \pm 3.5$ & 1.25\\
150625928 &      464 & fixed at $\Omega/2\pi = 1$ & $31.1 \pm 6.1$ & fixed at 1.0 & $1.30 \pm 0.36$ & $16.3 \pm 1.1$ & 1.38 & $33.7 \pm 11.$ & $6.00 \pm 0.29$ & $0.88 \pm 0.49$ & $16.4 \pm 3.8$ & 1.38\\
150625976 &      437 & fixed at $\Omega/2\pi = 1$ & $20.81 \pm 0.91$ & fixed at 1.0 & $2.59 \pm 0.14$ & $9.98 \pm 0.18$ & 1.43 & $17.69 \pm 0.76$ & $3.97 \pm 0.27$ & $2.85 \pm 0.20$ & $9.88 \pm 0.16$ & 1.44\\
150625984 &      297 & fixed at $\Omega/2\pi = 1$ & $28.5 \pm 4.9$ & fixed at 1.0 & $1.94 \pm 0.40$ & $5.32 \pm 0.42$ & 1.03 & $24.6 \pm 4.1$ & $4.88 \pm 0.62$ & $1.96 \pm 0.47$ & $5.27 \pm 0.27$ & 1.03\\
150625993 &      134 & fixed at $\Omega/2\pi = 1$ & $24.2 \pm 1.3$ & fixed at 1.0 & $2.17 \pm 0.17$ & $24.67 \pm 0.36$ & 1.66 & $27.1 \pm 2.9$ & $6.81 \pm 0.22$ & $1.37 \pm 0.23$ & $24.95 \pm 0.50$ & 1.62\\
150626063 &      258 & fixed at $\Omega/2\pi = 1$ & uncons & fixed at 1.0 & uncons & $15.3 \pm 7.0$ & 1.11 & uncons & $5.83 \pm 0.44$ & uncons & $15.3 \pm 6.1$ & 1.13\\
150626092 &      114 & fixed at $\Omega/2\pi = 1$ & $138.38 \pm 116.$ & fixed at 1.0 & uncons & $17.0 \pm 7.7$ & 1.44 & uncons & $6.39 \pm 0.18$ & uncons & $17.7 \pm 7.4$ & 1.38\\
150626109 &      408 & fixed at $\Omega/2\pi = 1$ & $31.3 \pm 4.8$ & fixed at 1.0 & $2.46 \pm 0.41$ & uncons & 1.10 & $28.2 \pm 5.2$ & $5.3 \pm 1.0$ & $2.22 \pm 0.58$ & $6.77 \pm 0.59$ & 1.11\\
150626124 &       57 & fixed at $\Omega/2\pi = 1$ & $21.7 \pm 1.3$ & fixed at 1.0 & $2.20 \pm 0.18$ & $31.77 \pm 0.54$ & 1.44 & $23.4 \pm 2.4$ & $6.25 \pm 0.22$ & $1.51 \pm 0.25$ & $32.01 \pm 0.81$ & 1.42\\
150626156 &      502 & fixed at $\Omega/2\pi = 1$ & $25.4 \pm 1.8$ & fixed at 1.0 & $3.01 \pm 0.41$ & $25.58 \pm 0.69$ & 1.29 & $36.6 \pm 9.4$ & $9.12 \pm 0.64$ & $1.12 \pm 0.50$ & $26.6 \pm 2.8$ & 1.28\\
150626163 &      296 & fixed at $\Omega/2\pi = 1$ & uncons & fixed at 1.0 & uncons & $11.2 \pm 4.5$ & 1.47 & uncons & $7.00 \pm 0.35$ & uncons & $11.4 \pm 3.2$ & 1.46\\
150626170 &      448 & fixed at $\Omega/2\pi = 1$ & $27.1 \pm 2.0$ & fixed at 1.0 & $2.35 \pm 0.27$ & $11.79 \pm 0.32$ & 1.39 & $27.5 \pm 3.2$ & $7.13 \pm 0.46$ & $1.71 \pm 0.32$ & $11.85 \pm 0.46$ & 1.39\\
150626177 &      361 & fixed at $\Omega/2\pi = 1$ & $33.5 \pm 10.$ & fixed at 1.0 & $1.44 \pm 0.67$ & $13.3 \pm 1.6$ & 1.31 & $37.3 \pm 20.$ & $8.50 \pm 0.72$ & $0.87 \pm 0.80$ & $13.5 \pm 4.9$ & 1.31\\
150626184 &      465 & fixed at $\Omega/2\pi = 1$ & $24.5 \pm 2.2$ & fixed at 1.0 & $2.56 \pm 0.30$ & $9.65 \pm 0.36$ & 1.19 & $20.8 \pm 1.8$ & $4.57 \pm 0.60$ & $2.70 \pm 0.38$ & $9.53 \pm 0.30$ & 1.18\\
150626192 &      409 & fixed at $\Omega/2\pi = 1$ & $40.6 \pm 15.$ & fixed at 1.0 & $1.56 \pm 0.70$ & $16.4 \pm 3.3$ & 1.44 & $33.3 \pm 9.1$ & $7.0 \pm 1.0$ & $1.56 \pm 0.68$ & $16.1 \pm 1.9$ & 1.46\\
150626222 &        8 & fixed at $\Omega/2\pi = 1$ & $30.0 \pm 2.4$ & fixed at 1.0 & $2.17 \pm 0.20$ & $33.90 \pm 0.93$ & 1.05 & $27.2 \pm 2.4$ & $5.28 \pm 0.39$ & $1.99 \pm 0.25$ & $33.6 \pm 1.0$ & 1.05\\
150626230 &      332 & fixed at $\Omega/2\pi = 1$ & uncons & fixed at 1.0 & uncons & $17.1 \pm 4.0$ & 1.13 & uncons & $6.64 \pm 0.62$ & uncons & $17.0 \pm 1.2$ & 1.13\\
150626238 &      257 & fixed at $\Omega/2\pi = 1$ & $33.9 \pm 5.4$ & fixed at 1.0 & $1.40 \pm 0.27$ & $8.81 \pm 0.44$ & 0.99 & $27.4 \pm 3.6$ & $4.58 \pm 0.28$ & $1.55 \pm 0.28$ & $8.68 \pm 0.36$ & 0.99\\
150626253 &       54 & fixed at $\Omega/2\pi = 1$ & $22.5 \pm 2.4$ & fixed at 1.0 & $2.45 \pm 0.43$ & $13.73 \pm 0.46$ & 1.20 & $22.7 \pm 3.7$ & $7.21 \pm 0.65$ & $1.80 \pm 0.54$ & $13.77 \pm 0.79$ & 1.20\\
150626260 &      112 & fixed at $\Omega/2\pi = 1$ & $25.2 \pm 2.5$ & fixed at 1.0 & $1.85 \pm 0.28$ & $16.09 \pm 0.49$ & 1.38 & $29.4 \pm 6.5$ & $7.01 \pm 0.29$ & $1.08 \pm 0.40$ & $16.2 \pm 1.4$ & 1.36\\
150626487 &       51 & fixed at $\Omega/2\pi = 1$ & $21.1 \pm 2.5$ & fixed at 1.0 & $1.55 \pm 0.26$ & $39.2 \pm 1.2$ & 1.31 & $20.6 \pm 3.3$ & $4.98 \pm 0.18$ & $1.33 \pm 0.34$ & $39.3 \pm 2.0$ & 1.32\\
150626494 &      133 & fixed at $\Omega/2\pi = 1$ & $15.09 \pm 0.87$ & fixed at 1.0 & $2.40 \pm 0.20$ & $19.72 \pm 0.26$ & 1.52 & $14.1 \pm 1.0$ & $4.24 \pm 0.22$ & $2.31 \pm 0.28$ & $19.70 \pm 0.33$ & 1.52\\
150626518 &      130 & fixed at $\Omega/2\pi = 1$ & uncons & fixed at 1.0 & uncons & $22.7 \pm 4.8$ & 1.42 & uncons & uncons & uncons & uncons & 1.35\\
150626553 &      276 & fixed at $\Omega/2\pi = 1$ & $20.8 \pm 1.5$ & fixed at 1.0 & $1.66 \pm 0.17$ & $66.63 \pm 0.96$ & 1.44 & $20.3 \pm 1.9$ & $5.08 \pm 0.12$ & $1.42 \pm 0.22$ & $66.7 \pm 1.3$ & 1.46\\
150626564 &      153 & fixed at $\Omega/2\pi = 1$ & $11.0 \pm 1.1$ & fixed at 1.0 & $2.71 \pm 0.53$ & $32.37 \pm 0.91$ & 1.57 & $11.2 \pm 2.3$ & $5.19 \pm 0.43$ & $2.06 \pm 0.98$ & $32.3 \pm 2.9$ & 1.58\\
150626577 &      504 & fixed at $\Omega/2\pi = 1$ & uncons & fixed at 1.0 & uncons & $66.5 \pm 27.$ & 1.57 & uncons & $4.84 \pm 0.19$ & uncons & $66.5 \pm 16.$ & 1.59\\
150626593 &      137 & fixed at $\Omega/2\pi = 1$ & $21.6 \pm 4.7$ & fixed at 1.0 & $1.51 \pm 0.47$ & $58.9 \pm 5.3$ & 1.14 & $21.9 \pm 6.7$ & $4.72 \pm 0.32$ & $1.23 \pm 0.60$ & $59.0 \pm 7.0$ & 1.14\\
150626620 &      483 & fixed at $\Omega/2\pi = 1$ & uncons & fixed at 1.0 & uncons & $41.0 \pm 15.$ & 1.24 & uncons & $3.86 \pm 0.28$ & uncons & $41.0 \pm 19.$ & 1.24\\
150626627 &      599 & fixed at $\Omega/2\pi = 1$ & $41.6 \pm 9.2$ & fixed at 1.0 & $0.60 \pm 0.20$ & $35.2 \pm 2.9$ & 1.45 & $31.9 \pm 5.4$ & $3.726 \pm 0.079$ & $0.78 \pm 0.20$ & $35.0 \pm 1.6$ & 1.51\\
150626636 &      505 & fixed at $\Omega/2\pi = 1$ & $33.4 \pm 31.$ & fixed at 1.0 & uncons & $42.1 \pm 18.$ & 1.23 & uncons & $3.26 \pm 0.25$ & uncons & $42.1 \pm 15.$ & 1.24\\
150626643 &      501 & fixed at $\Omega/2\pi = 1$ & $22.4 \pm 4.1$ & fixed at 1.0 & $1.22 \pm 0.31$ & $38.3 \pm 2.8$ & 1.47 & $18.3 \pm 2.7$ & $3.69 \pm 0.18$ & $1.49 \pm 0.33$ & $38.2 \pm 1.3$ & 1.47\\
150626652 &      110 & fixed at $\Omega/2\pi = 1$ & uncons & fixed at 1.0 & uncons & $56.1 \pm 17.$ & 1.12 & uncons & $5.09 \pm 0.20$ & uncons & $56.8 \pm 15.$ & 1.13\\
150626685 &      399 & fixed at $\Omega/2\pi = 1$ & $14.0 \pm 1.3$ & fixed at 1.0 & $2.32 \pm 0.31$ & $72.7 \pm 1.8$ & 1.31 & $13.2 \pm 1.6$ & $4.46 \pm 0.25$ & $2.19 \pm 0.45$ & $72.7 \pm 2.6$ & 1.31\\
150626724 &       40 & fixed at $\Omega/2\pi = 1$ & $16.8 \pm 8.5$ & fixed at 1.0 & $2.3 \pm 1.5$ & $76.9 \pm 21.$ & 1.36 & $14.7 \pm 7.1$ & $4.6 \pm 1.6$ & $2.5 \pm 2.0$ & $77.2 \pm 26.$ & 1.36\\
150626751 &       54 & fixed at $\Omega/2\pi = 1$ & $21.8 \pm 15.$ & fixed at 1.0 & $1.3 \pm 1.2$ & $63.8 \pm 29.$ & 1.14 & uncons & $4.52 \pm 0.58$ & uncons & $63.9 \pm 28.$ & 1.14\\
150626766 &      108 & fixed at $\Omega/2\pi = 1$ & $76.5 \pm 37.$ & fixed at 1.0 & $0.54 \pm 0.36$ & $23.3 \pm 9.1$ & 2.01 & $44.9 \pm 6.8$ & $3.92 \pm 0.19$ & $0.88 \pm 0.24$ & $21.6 \pm 1.5$ & 2.10\\
150626773 &      434 & fixed at $\Omega/2\pi = 1$ & $23.3 \pm 1.4$ & fixed at 1.0 & $1.58 \pm 0.12$ & $24.26 \pm 0.39$ & 1.66 & $18.80 \pm 0.96$ & $3.67 \pm 0.12$ & $1.91 \pm 0.13$ & $24.02 \pm 0.26$ & 1.68\\
150626781 &      500 & fixed at $\Omega/2\pi = 1$ & $23.3 \pm 1.9$ & fixed at 1.0 & $1.59 \pm 0.16$ & $20.20 \pm 0.28$ & 1.47 & $19.8 \pm 1.4$ & $4.05 \pm 0.13$ & $1.77 \pm 0.18$ & $20.09 \pm 0.33$ & 1.49\\
150626788 &       36 & fixed at $\Omega/2\pi = 1$ & uncons & fixed at 1.0 & uncons & $40.8 \pm 8.2$ & 1.01 & uncons & $9.4 \pm 2.9$ & uncons & $41.0 \pm 3.8$ & 1.01\\
150626822 &      514 & fixed at $\Omega/2\pi = 1$ & $16.5 \pm 1.5$ & fixed at 1.0 & $2.87 \pm 0.38$ & $14.94 \pm 0.52$ & 1.27 & $14.2 \pm 1.4$ & $4.20 \pm 0.58$ & $3.14 \pm 0.58$ & $14.81 \pm 0.49$ & 1.27\\
150626829 &      317 & fixed at $\Omega/2\pi = 1$ & $14.4 \pm 1.1$ & fixed at 1.0 & $3.11 \pm 0.49$ & $15.18 \pm 0.38$ & 1.12 & $14.1 \pm 2.4$ & $6.37 \pm 0.60$ & $2.39 \pm 0.77$ & $15.0 \pm 1.1$ & 1.13\\
150626844 &      505 & fixed at $\Omega/2\pi = 1$ & $21.3 \pm 1.3$ & fixed at 1.0 & $2.30 \pm 0.18$ & $13.71 \pm 0.24$ & 1.37 & $18.7 \pm 1.2$ & $4.75 \pm 0.27$ & $2.33 \pm 0.24$ & $13.63 \pm 0.26$ & 1.38\\
150626851 &       93 & fixed at $\Omega/2\pi = 1$ & $21.7 \pm 1.6$ & fixed at 1.0 & $2.54 \pm 0.28$ & $14.53 \pm 0.37$ & 1.06 & $21.1 \pm 2.2$ & $5.99 \pm 0.46$ & $2.07 \pm 0.37$ & $14.54 \pm 0.44$ & 1.06\\
150626888 &      726 & fixed at $\Omega/2\pi = 1$ & $27.0 \pm 2.7$ & fixed at 1.0 & $2.58 \pm 0.31$ & $6.64 \pm 0.51$ & 1.12 & $21.9 \pm 1.8$ & $3.68 \pm 0.75$ & $2.98 \pm 0.38$ & $6.54 \pm 0.24$ & 1.10\\
150626896 &      593 & fixed at $\Omega/2\pi = 1$ & $34.7 \pm 4.9$ & fixed at 1.0 & $1.92 \pm 0.29$ & $7.0 \pm 1.3$ & 1.17 & $25.2 \pm 2.3$ & $3.46 \pm 0.59$ & $2.45 \pm 0.29$ & $6.77 \pm 0.26$ & 1.17\\
150626903 &      462 & fixed at $\Omega/2\pi = 1$ & $54.1 \pm 15.$ & fixed at 1.0 & $1.60 \pm 0.42$ & $6.1 \pm 2.2$ & 0.87 & $43.5 \pm 9.1$ & $4.64 \pm 0.68$ & $1.63 \pm 0.37$ & $5.99 \pm 0.51$ & 0.88\\
150626910 &       65 & fixed at $\Omega/2\pi = 1$ & $31.1 \pm 1.4$ & fixed at 1.0 & $2.24 \pm 0.11$ & $17.82 \pm 0.85$ & 1.32 & $24.35 \pm 0.91$ & $3.62 \pm 0.28$ & $2.63 \pm 0.13$ & $17.35 \pm 0.28$ & 1.34\\
150626918 &       70 & fixed at $\Omega/2\pi = 1$ & uncons & fixed at 1.0 & uncons & $31.2 \pm 10.$ & 1.05 & uncons & $8.32 \pm 0.36$ & uncons & $30.6 \pm 11.$ & 0.99\\
150626957 &      378 & fixed at $\Omega/2\pi = 1$ & $46.4 \pm 13.$ & fixed at 1.0 & $1.20 \pm 0.41$ & $7.0 \pm 2.9$ & 1.32 & $30.4 \pm 4.0$ & $2.06 \pm 0.86$ & $1.85 \pm 0.26$ & $6.97 \pm 0.73$ & 1.30\\
150626979 &      424 & fixed at $\Omega/2\pi = 1$ & $80.3 \pm 49.$ & fixed at 1.0 & $0.78 \pm 0.67$ & $7.5 \pm 2.9$ & 0.94 & uncons & $6.53 \pm 0.43$ & uncons & $7.6 \pm 3.4$ & 0.94\\
150626989 &      251 & fixed at $\Omega/2\pi = 1$ & uncons & fixed at 1.0 & uncons & $3.97 \pm 0.32$ & 1.11 & uncons & $5.47 \pm 0.63$ & uncons & $3.8 \pm 1.2$ & 1.11\\
150627023 &      559 & fixed at $\Omega/2\pi = 1$ & $26.4 \pm 1.4$ & fixed at 1.0 & $2.32 \pm 0.16$ & $13.17 \pm 0.25$ & 1.71 & $23.0 \pm 1.3$ & $5.00 \pm 0.31$ & $2.30 \pm 0.20$ & $13.04 \pm 0.26$ & 1.73\\
150627030 &      491 & fixed at $\Omega/2\pi = 1$ & $35.7 \pm 3.1$ & fixed at 1.0 & $2.26 \pm 0.20$ & $9.1 \pm 3.8$ & 1.09 & $26.5 \pm 1.5$ & $3.07 \pm 0.61$ & $2.79 \pm 0.21$ & $8.79 \pm 0.33$ & 1.10\\
150627038 &      348 & fixed at $\Omega/2\pi = 1$ & $32.6 \pm 4.6$ & fixed at 1.0 & $2.97 \pm 0.53$ & $6.9 \pm 2.1$ & 1.26 & $26.5 \pm 3.1$ & $3.8 \pm 1.7$ & $3.20 \pm 0.62$ & $6.77 \pm 0.45$ & 1.26\\
150627047 &      506 & fixed at $\Omega/2\pi = 1$ & $44.7 \pm 6.6$ & fixed at 1.0 & $2.13 \pm 0.32$ & $6.92 \pm 0.38$ & 1.66 & $34.1 \pm 3.7$ & $4.49 \pm 0.78$ & $2.35 \pm 0.32$ & $6.60 \pm 0.34$ & 1.65\\
150627054 &      217 & fixed at $\Omega/2\pi = 1$ & $49.9 \pm 9.4$ & fixed at 1.0 & $2.13 \pm 0.41$ & $9.86 \pm 0.77$ & 1.28 & $41.6 \pm 7.4$ & $6.3 \pm 1.1$ & $1.96 \pm 0.44$ & $9.50 \pm 0.66$ & 1.28\\
150627082 &      408 & fixed at $\Omega/2\pi = 1$ & $28.0 \pm 4.3$ & fixed at 1.0 & $2.50 \pm 0.56$ & $7.78 \pm 0.48$ & 1.33 & $29.3 \pm 7.3$ & $7.2 \pm 1.0$ & $1.71 \pm 0.69$ & $7.84 \pm 0.78$ & 1.32\\
150627091 &      377 & fixed at $\Omega/2\pi = 1$ & $47.9 \pm 16.$ & fixed at 1.0 & $1.29 \pm 0.59$ & $6.6 \pm 1.1$ & 1.55 & $58.7 \pm 50.$ & $8.51 \pm 0.66$ & uncons & $6.8 \pm 3.1$ & 1.54\\
150627162 &      192 & fixed at $\Omega/2\pi = 1$ & $42.7 \pm 3.0$ & fixed at 1.0 & $1.90 \pm 0.13$ & $9.06 \pm 0.97$ & 1.15 & $31.9 \pm 1.5$ & $3.61 \pm 0.32$ & $2.24 \pm 0.13$ & $8.70 \pm 0.19$ & 1.13\\
150627978 &       49 & fixed at $\Omega/2\pi = 1$ & $36.5 \pm 4.7$ & fixed at 1.0 & $2.42 \pm 0.31$ & uncons & 1.16 & $26.9 \pm 2.4$ & $3.32 \pm 0.96$ & $2.98 \pm 0.38$ & $5.03 \pm 0.26$ & 1.18\\
\enddata
\tablenotetext{a}{REFLECT*CompTT model}
\tablenotetext{b}{(10-1000 keV)}
\tablenotetext{c}{CompTT model}

\label{tbl:spec}
\end{deluxetable*}

\end{center}

\end{document}